\documentclass[11pt]{article}
\usepackage{amsmath,amssymb,mathtools}
\usepackage{graphicx}
\usepackage[sc,small]{caption}
\usepackage[retainorgcmds]{IEEEtrantools}
\usepackage{dsfont}
\usepackage{url}
\usepackage{jheppub}

\newcommand{\be}{\begin{eqnarray}}
\newcommand{\ee}{\end{eqnarray}}

\newcommand{\beqn}{\begin{eqnarray}}
\newcommand{\eeqn}{\end{eqnarray}}

\begin{document}

%\hfill   1112.xxxx \\

\hfill LMU-ASC 45/20\\
\vspace{1cm}

\begin{center}
{\bf\LARGE
Closed string disk amplitudes in the pure spinor formalism \\  }

\vspace{2.5cm}

{\large
{\bf Andreas Bischof and Michael Haack}
\vspace{1cm}

{\it 
Arnold Sommerfeld Center for Theoretical Physics \\ 
Ludwig-Maximilians-Universit\"at M\"unchen \\ 
Theresienstrasse 37, 80333 M\"unchen, Germany
}
}

\end{center}
\vspace{8mm}

%\abstract
\begin{center}
{\bf Abstract}\\
\end{center}
We evaluate closed string disk amplitudes in the pure spinor formalism. We focus on low point functions (two- and one-point functions) but our analysis is also relevant for higher $n$-point functions. Amongst others we discuss issues arising due to the gauge fixing of the conformal Killing group of the disk and due to the zero mode prescription in the pure spinor formalism. As expected, in the end we find agreement with the known results from the RNS formalism. 
\clearpage
 \tableofcontents

%%%%%%%%%%%%%%%%%%%%%%%%%%%%%%%%%%%%%%

\section{Introduction}
\label{intro}

The pure spinor formalism \cite{Berkovits:2000fe} has become a real contender for computations in string perturbation theory. It has some noticeable advantages over the more traditional approaches based on the RNS and the Green-Schwarz formalisms, given that it is manifestly space-time supersymmetric and Lorentz covariant. As the formalism does not make use of any world-sheet spinors, one does not have to sum over spin structures. This is a significant simplification in calculations of higher loop amplitudes compared to the RNS formalism. Thus it should not come as a surprise that certain calculations were first (or even only) performed using the pure spinor formalism. This includes the computation of the complete quartic effective action of type II string theory at sphere level (including the RR fields of the RNS formalism) \cite{Policastro:2006vt}, the calculation of an arbitrary $n$-point amplitude of massless open strings on the disk in type I string theory \cite{Mafra:2011nv} and the computation of the closed string four-point {\it 3-loop} amplitude in type II string theory at low energy \cite{Gomez:2013sla}. 

However, the calculation of purely closed string amplitudes at the disk level seems to be lacking in the literature on the pure spinor formalism and it is the purpose of this note to fill this small gap.\footnote{Of course, these amplitudes could be obtained indirectly from purely open strings on the disk, applying the relations found in \cite{Stieberger:2009hq} in the context of the RNS formalism. What we mean here is a direct calculation of the closed string disk amplitudes in the pure spinor formalism.} The case of one closed and two open strings was discussed in \cite{Alencar:2008fy, Alencar:2011tk}. However, the case with only closed string vertex operators on the disk presents new questions, given that fixing the conformal Killing group (CKG) of the disk only allows to fix one and a half closed string vertex operators. Moreover, the disk amplitude with a single closed string vertex operator would naively vanish with the usual tree level prescription of dealing with the fermionic zero modes in the pure spinor formalism. Both of these issues will be addressed in the following. The answers are already hidden in the literature and we are going to collect and apply these known results to the mentioned closed string amplitudes on the disk. More concretely, we are going to calculate a two-point function of closed string states which would correspond to massless NSNS states in the RNS formalism (i.e.\ the graviton, dilaton and Kalb-Ramond field) by following the gauge fixing procedure described in \cite{Hoogeveen:2007tu,Hoogeveen:2010yfa}. Afterwards we calculate the disk one-point function of the same states using the alternative zero mode prescription introduced in \cite{Berkovits:2016xnb}.\footnote{A different approach to calculating low point functions, using the usual zero mode prescription, was employed in \cite{Kashyap:2020tgx}, which deals with the open string two-point function on the disk. It would be interesting to better understand the relation between the two methods.}

Needless to say that the results agree with the corresponding calculations performed in the RNS formalism. In that formalism the closed string two-point function on the disk in superstring theory was calculated in \cite{Klebanov:1995ni,Gubser:1996wt,Garousi:1996ad,Hashimoto:1996kf,Hashimoto:1996bf} (see also \cite{Becker:2011bw,Aldi:2020dvw} which revisited the topic more recently). The dilaton one-point function, on the other hand, was calculated for the bosonic string in \cite{Douglas:1986eu,Liu:1987nz} and a generalization to the superstring was performed in \cite{Ohta:1987nq}.

%%%%%%%%%%%%%%%%%%%%%%%%%%%%%%%%%%%%%%

\section{The pure spinor formalism}
\label{PSF}

Let us begin with a short introduction to those aspects of the pure spinor formalism that are relevant for our question. 

%%%%%%%%%%%%%%%%%%%%%%%%%%%%%%%%%%%%%%

\subsection{Matter and ghost CFT of the pure spinor formalism}\label{sec::pdf_matter}

The action of the pure spinor formalism is given by\footnote{See appendix \ref{app:notation} for our conventions and notation.}
\begin{IEEEeqnarray}{rCl}
S=\frac{1}{2\pi}\int\mathrm d^2z\,\left(\frac12\partial X^m\overline\partial X_m+p_\alpha\overline\partial\theta^\alpha+\overline p_\alpha\partial\overline\theta^\alpha - w_\alpha\overline\partial\lambda^\alpha-\overline w_\alpha\partial\overline\lambda^\alpha\right) .\label{eq::psfaction2}
\end{IEEEeqnarray}
It leads to the holomorphic energy momentum tensor
\begin{IEEEeqnarray}{rCl}
T(z)=-\frac12\partial X^m\partial X_m-p_\alpha\partial\theta^\alpha+w_\alpha\partial\lambda^\alpha
\label{eq::psfemt}
\end{IEEEeqnarray}
and a similar expression for the anti-holomorphic energy momentum tensor.\footnote{Of course, all the formulas in the rest of this subsection have an obvious  antiholomorphic counterpart.} The theory has a vanishing central charge in ten spacetime dimensions \cite{Berkovits:2000fe}.

It is convenient to introduce the supersymmetric fields
\begin{IEEEeqnarray}{rCl}
\Pi^m&=&\partial X^m+\frac 1 2 (\theta\gamma^m\partial\theta)\ , \label{Pi_def} \\
d_\alpha&=&p_\alpha-\frac 1 2 \left(\partial X^m+\frac 1 4 (\theta\gamma^m\partial\theta)\right)(\gamma_m\theta)_\alpha\ , \label{dalpha_def}
\end{IEEEeqnarray}
because these conformal primaries of weight $h=1$ appear in the vertex operators of massless fields and, thus, play an important role in the calculation of scattering amplitudes in the pure spinor formalism, as we will review below. We will need the following OPEs:
\begin{IEEEeqnarray}{C}
\begin{IEEEeqnarraybox}[][c]{rClCrCl}
\IEEEstrut
X^m(z,\overline z)X^n(w,\overline w)&=&-\eta^{mn}\ln{|z-w|^2}\ , & \qquad & p_\alpha(z)\theta^\beta(w) & = & \frac{\delta_\alpha^{\hphantom\alpha\beta}}{z-w}\ , \\
\Pi^m(z)\Pi^n(w) & = & \frac{-\eta^{mn}}{(z-w)^2}\ , & \quad & d_\alpha(z) d_\beta(w) & = & -\frac{\gamma^m_{\alpha\beta}\Pi_m(w)}{z-w}\ , \\
d_\alpha(z)\Pi^m(w) & = & \frac{(\gamma^m\partial\theta)_{\alpha}(w)}{z-w}\ , & \quad & d_\alpha(z)\theta^\beta(w) & = & \frac{\delta^{\hphantom\alpha\beta}_\alpha}{z-w}\ .
\IEEEstrut
\end{IEEEeqnarraybox}\label{psfOPE}
\end{IEEEeqnarray}

In the pure spinor formalism $\lambda^{\alpha}$ is a commuting $SO(1,9)$ Weyl spinor. Therefore, it contributes to the Lorentz current and the total contribution from the spacetime fermions is given by
\begin{IEEEeqnarray}{rCl}
M^{mn} = \frac 1 2 (p\gamma^{mn}\theta) - \frac 1 2 (w \gamma^{mn} \lambda) \equiv \Sigma^{mn} - N^{mn} \ . 
\label{eq::lorentz_current}
\end{IEEEeqnarray} 
The relative sign between the two fermionic contributions can be traced back to the relative sign in the action \eqref{eq::psfaction2}. The Lorentz current \eqref{eq::lorentz_current} is determined by demanding
\be
\delta \Psi^\alpha =  \frac{\epsilon_{mn}}{2} {\rm Res}_{z=w} \Big( M^{mn}(z)  \Psi^\alpha (w) \Big) = - \frac{\epsilon_{mn}}{4} (\gamma^{mn})^{\alpha}\, \! _\beta \Psi^\beta(w)\ , \qquad \Psi^\alpha \in \{ \theta^\alpha, \lambda^\alpha \}\ .
\label{deltaPsi}
\ee
Note that the sign is in line with 
\begin{IEEEeqnarray}{rCl}
\delta \psi^k = \frac{\epsilon_{mn}}{2} {\rm Res}_{z=w} \Big( \psi^m \psi^n (z) \psi^k (w) \Big) =  - \epsilon^k \,\! _m \psi^m(w)
\end{IEEEeqnarray}
for the Lorentz current $\psi^m \psi^n$ of the world-sheet fermions in the RNS formalism. Equation \eqref{deltaPsi} implies the OPE
\begin{IEEEeqnarray}{rCL}
N^{mn}(z)\lambda^\alpha(w)&=&\frac{(\gamma^{mn})^{\alpha}_{\hphantom{\alpha}\beta}\lambda^{\beta}(w)}{2(z-w)}\ .
\label{ghostcurrent2}
\end{IEEEeqnarray}
Moreover, the OPEs of two $N$s and two $M$s are given by\footnote{Formula \eqref{ghostcurrent1} has a sign for the simple pole terms which differs from much of the literature. However, it is consistent with our conventions. Moreover, it is the full $M^{mn}$ which appears in the vertex operator \eqref{V1} below, if one combines the contributions from the 3rd and 4th term in \eqref{V1}.}
\begin{IEEEeqnarray}{rCl}
N^{mn}(z)N^{pq}(w)&=& - \frac{\eta^{p[n}N^{m]q}(w)-\eta^{q[n}N^{m]p}(w)}{z-w}-3\frac{\eta^{m[q}\eta^{p]n}}{(z-w)^2}\ , 
\label{ghostcurrent1} \\
M^{mn}(z)M^{pq}(w)&=&\frac{\eta^{p[n}M^{m]q}(w)-\eta^{q[n}M^{m]p}(w)}{z-w}+\frac{\eta^{m[q}\eta^{p]n}}{(z-w)^2}\ .
\label{eq::Lorentz_current}
\end{IEEEeqnarray}

Finally, let us remind the reader that nilpotency of the BRST operator
\begin{IEEEeqnarray}{C}
Q=\oint\frac{\mathrm{d}z}{2\pi i}\,\lambda^\alpha(z) d_\alpha(z)\ ,
\label{BRST}
\end{IEEEeqnarray}
i.e.\ $Q^2=0$, implies that $\lambda$ has to be a pure spinor, i.e.
\begin{IEEEeqnarray}{rCl}
(\lambda\gamma^m\lambda)=0\ ,
\label{purespinor}
\end{IEEEeqnarray}
and similarly for the right-movers. 

%%%%%%%%%%%%%%%%%%%%%%%%%%%%%%%%%%%%%%

\subsection{Massless vertex operators in the pure spinor formalism}
\label{sec_Vops}

In this article we are interested in the scattering amplitudes of closed strings on the disk (we actually always mean the upper half plane $\mathbb H_+$ when talking about the disk). More precisely, we focus on those massless states whose polarization tensor can be obtained via a tensor product of two vectors, i.e.\ $\epsilon_{mn} = \xi_m \otimes \overline \xi_n$. In the RNS formalism, these correspond to the NSNS states, i.e.\ the graviton, the Kalb-Ramond field and the dilaton. The vertex operator for such a state is given by 
\begin{IEEEeqnarray}{C}
V^{(a,b)}(z, \overline z) =  V^{(a)}(z) \otimes \overline V \, \! ^{(b)}(\overline z)\ , \qquad a,b \in \{ 0,1 \}\ ,
\label{closed_vops}
\end{IEEEeqnarray}
where \cite{Berkovits:2000fe}
\begin{IEEEeqnarray}{rCl}
V^{(0)}(z)&=&\left[\lambda^\alpha A_\alpha(X,\theta)\right](z)\ , \label{V0} \\
V^{(1)}(z)&=&\left[\partial\theta^\alpha A_\alpha(X,\theta)+\Pi^m A_m(X,\theta)+d_\alpha W^\alpha(X,\theta)+\frac12N^{mn}\mathcal F_{mn}(X,\theta)\right](z)
\label{V1}
\IEEEeqnarraynumspace
\end{IEEEeqnarray}
are related to the massless open string vertex operators.\footnote{In \eqref{closed_vops} we made the fact explicit that the product of the open string vertex operators involves a tensor product of the two polarization vectors. In the following we will omit the $\otimes$ symbol and leave it implicit. Moreover, in the literature $V^{(0)}$ and $V^{(1)}$ are often denoted by $V$ and $U$, respectively.} In \eqref{V0} and \eqref{V1}, $A_\alpha, A_m, W^\alpha$ and $\mathcal F_{mn}$ are all spacetime superfields (the superfields of super-Maxwell theory). The first vertex operator $V^{(0)}$ is BRST closed, i.e.\
\begin{IEEEeqnarray}{rCl}
Q V^{(0)} = 0\ ,
\label{QV0}
\end{IEEEeqnarray}
and the second vertex operator $V^{(1)}$ fulfils 
\begin{IEEEeqnarray}{rCl}
QV^{(1)}=\partial V^{(0)}\ .
\label{QV1}
\end{IEEEeqnarray}
Hence it is in the BRST cohomology once we integrate it over the world-sheet. Consequently, $V^{(0)}$ and $V^{(1)}$ are called the unintegrated and integrated vertex operator, respectively. Analogous statements hold for the right-moving part of \eqref{closed_vops}. 

The fields $A_m, W^\alpha$ and $\mathcal F_{mn}$ in \eqref{V1} are not independent. Rather, they are the field strengths given by \cite{Berkovits:2002zk}
\begin{IEEEeqnarray}{rCl}
A_m & = & \frac18\gamma^{\alpha\beta}_mD_\alpha A_\beta \ ,\\
W^\alpha & = & \frac1{10}\gamma_m^{\alpha\beta}(D_\beta A^m-\partial^mA_\beta)\ ,\\
\mathcal F_{mn} & = & \partial_mA_n-\partial_nA_m\ ,
\end{IEEEeqnarray}
where we introduced the superderivative
\begin{IEEEeqnarray}{rCl}
D_\alpha=\frac{\partial}{\partial\theta^\alpha}+\frac 1 2(\gamma^m\theta)_\alpha\partial_m\ .\label{canonicalderivative}
\end{IEEEeqnarray}
The superfields fulfil the following equations \cite{Berkovits:2002zk,Witten:1985nt}
\begin{IEEEeqnarray}{l}
\begin{IEEEeqnarraybox}[][c]{rClCrCl}
\IEEEstrut
2 D_{(\alpha} A_{\beta)} & = & \gamma^m_{\alpha \beta} A_m\ , & \quad & D_\alpha A_m & = & (\gamma_m W)_\alpha + \partial_m A_\alpha\ , \\
D_\alpha {\cal F}_{mn} & = & 2 \partial_{[m} ( \gamma_{n]} W)_\alpha\ , & \quad & D_\alpha W^\beta & = & \frac14 (\gamma^{mn})_\alpha \, \! ^\beta  {\cal F}_{mn} \ .
\IEEEstrut
\end{IEEEeqnarraybox}
\label{eq::eoms_superfields}
\end{IEEEeqnarray}

When calculating amplitudes one needs the $\theta$-expansion of the superfields. Restricting to the bosonic spacetime degrees of freedom (of a vector field with polarization vector $\xi_m$), relevant for the concrete calculations in the later sections, and making the gauge choice $\theta^\alpha A_\alpha(X,\theta)=0$, the expansions can be found, for instance, in (10.2.31) of \cite{Schlotterer:2011psa}:\footnote{The expansion in \cite{Schlotterer:2011psa} is more general, allowing to describe also fermionic spacetime degrees of freedom. Moreover, note that our momenta are real, i.e.\ they differ from the corresponding momenta of \cite{Schlotterer:2011psa} by a factor of $i$.}
\begin{IEEEeqnarray}{rCl}
A_{\alpha}(X,\theta)&=&e^{ik\cdot X}\left\{\frac{\xi_m}{2}(\gamma^m\theta)_\alpha-\frac{1}{16}(\gamma_p\theta)_\alpha(\theta\gamma^{mnp}\theta)i k_{[m}\xi_{n]} +\mathcal O(\theta^5)\right\}, \nonumber \\
A_m(X,\theta)&=&e^{ik\cdot X}\left\{\xi_m-\frac14 ik_p(\theta\gamma_m^{\hphantom{m}pq}\theta)\xi_q +\mathcal O(\theta^4)\right\} , \nonumber \\
W^\alpha(X,\theta)&=&e^{ik\cdot X}\left\{-\frac12ik_{[m}\xi_{n]}(\gamma^{mn}\theta)^\alpha +\mathcal O(\theta^3)\right\}, \nonumber \\
\mathcal{F}_{mn}(X,\theta)&=&e^{ik\cdot X}\left\{2ik_{[m}\xi_{n]}-\frac12ik_{[p}\xi_{q]}ik_{[m}(\theta\gamma_{n]}^{\hphantom{n]}pq}\theta) +\mathcal O(\theta^4)\right\}. 
\label{expansion}
\end{IEEEeqnarray}
We only displayed the expansions up to the order in $\theta$ that is relevant for our purposes. Moreover, we organized the $X^m$-dependence of the superfields into plane waves with momentum $k^m$. Note that all the superfields in \eqref{expansion} depend holomorphically on $z$, i.e.\ $X^m = X^m(z)$. This means that we use the separation of $X^m(z, \overline z)$ into left- and right-movers, i.e.\ $X^m(z, \overline z) = X^m(z)  + \overline X^m(\overline z)$. $\overline V \, \! ^{(b)}$ in \eqref{closed_vops} is obtained from \eqref{V0}, \eqref{V1} and \eqref{expansion} by replacing the left moving fields $X(z), \theta(z), \lambda^\alpha (z)$ with their right moving counterparts $\overline X(\overline z), \overline \theta(\overline z), \overline \lambda^\alpha (\overline z)$ and $\xi_m$ with $\overline \xi_m$. In this way the full closed string vertex operator \eqref{closed_vops} contains a factor $e^{i k \cdot X (z, \overline z)}$.\footnote{A side comment: If one wanted to make the relation between the closed string vertex operator \eqref{closed_vops} with two open string vertex operators precise, one would have to take into account that the closed string momentum is split over the two open string vertex operators. This is well known from the RNS formalism, cf.\ \cite{Garousi:1996ad,Hashimoto:1996bf}, and in the context of the pure spinor formalism it was discussed in \cite{Alencar:2008fy}.}

Note that we allow $a \neq b$ in \eqref{closed_vops}. When calculating closed string amplitudes on the sphere, the corresponding conformal Killing group allows to fix three closed string vertex operators, leaving all the others integrated. Thus, in that case the choice $a=b$ is possible and always made in the literature. However, the conformal Killing group of the disk does not allow to fix the positions of two closed string vertex operators completely. Thus, when calculating a disk amplitude with only closed strings, one has to allow the possibility $a \neq b$ in \eqref{closed_vops}. We will see this more concretely in section \ref{2pt}. This possibility was also discussed in \cite{Alencar:2011tk,Grassi:2004ih}.

In sections \ref{2pt} and \ref{1pt} we will calculate two- and one-point functions of closed strings on this disk. For that purpose it is useful to rewrite the right-moving part of the vertex operator \eqref{closed_vops} using the doubling trick, in order to allow for a unified treatment of the left- and right-movers. This is done in the rest of this subsection. 

Concretely, in the following we consider a type II theory in a flat ten dimensional spacetime, which contains a D$p$-brane that is spanned in the $X_1\times X_2\times\ldots\times X_p$ plane. As usual, we use the fact that the D-brane is infinitely heavy in the small coupling regime, i.e.\ it can absorb an arbitrary amount of momentum in the $X_{p+1},\ldots,X_9$ directions transversal to the D-brane. Thus, momentum is effectively only conserved along the D-brane in the perturbative regime that we are working in. 

Left- and rightmovers separately have the standard correlators on the upper half plane 
\begin{IEEEeqnarray}{l}
\begin{IEEEeqnarraybox}[][c]{rCl}
\IEEEstrut
\langle X^m(z)X^n(w)\rangle&=&-\eta^{mn}\ln(z-w)\ ,\\
\langle p_{\alpha}(z)\theta^\beta(w)\rangle&=&\frac{\delta_\alpha^{\hphantom{\alpha}\beta}}{z-w}\ ,\\
\langle w_{\alpha}(z)\lambda^\beta(w)\rangle&=&\frac{\delta_\alpha^{\hphantom{\alpha}\beta}}{z-w}\ ,
\IEEEstrut
\end{IEEEeqnarraybox}
\label{eq::correlator}
\end{IEEEeqnarray}
where the antiholomorphic part is analogous. At the boundary of $\mathbb H_+$, i.e.\ at the real axis, the first $p+1$ components of the world-sheet fields satisfy Neumann  boundary conditions and the remaining $9-p$ components Dirichlet boundary conditions. Both of these boundary conditions impose non-vanishing correlators between the holomorphic and antiholomorphic parts of the fields. We can simplify the calculations by employing the doubling trick, i.e.\ we replace the right moving spacetime vectors and spacetime spinors by
\begin{IEEEeqnarray}{rCl}
\text{vectors: }\overline X^m(\overline z)=D^m_{\hphantom mn}X^n(\overline z)\ , \quad\text{spinors: }\overline\Psi^\alpha(\overline z) = M^\alpha_{\hphantom{\alpha}\beta}\Psi^\beta(\overline z) \quad \text{ or } \quad \overline{\Psi}_\alpha(\overline z) = N_\alpha^{\hphantom{\alpha}\beta}\Psi_\beta(\overline z)\ , \nonumber  \\
\label{eq::replace}
\end{IEEEeqnarray}
with $\Psi^\alpha \in \{ \theta^\alpha, \lambda^\alpha\}$ and $\Psi_\alpha \in \{ p_\alpha, w_\alpha\}$ and constant matrices $D,M$ and $N$. This corresponds to extending the world-sheet fields to the entire complex plane and allows us to use only the correlators in \eqref{eq::correlator}, leading to
\begin{IEEEeqnarray}{l}
\begin{IEEEeqnarraybox}[][c]{rClCrCl}
\IEEEstrut
\langle X^m(z)\overline X^n(\overline w)\rangle&=&-D^{mn}\ln(z-\overline w)\ ,\\
\langle p_{\alpha}(z)\overline\theta^\beta(\overline w)\rangle&=&\frac{M^{\beta}_{\hphantom{\beta} \alpha}}{z-\overline w}\ ,&\quad&\langle \overline p_{\alpha}(\overline z)\theta^\beta(w)\rangle&=&\frac{N_\alpha^{\hphantom{\alpha}\beta}}{\overline z-w}\ ,\\
\langle w_\alpha(z)\overline \lambda^\beta(\overline w)\rangle&=&\frac{M^{\beta}_{\hphantom{\beta} \alpha}}{z-\overline w}\ ,&\quad&\langle \overline w_\alpha(\overline z)\lambda^\beta(w)\rangle&=&\frac{N_\alpha^{\hphantom\alpha\beta}}{\overline z-w}\ .
\IEEEstrut
\end{IEEEeqnarraybox}
\label{eq::correlator2ptdisk}
\end{IEEEeqnarray}
The matrix $D^{mn}$ is the same as in the RNS formalism \cite{Garousi:1996ad,Hashimoto:1996bf}. Concretely, $D^{mn}$ is given by
\begin{IEEEeqnarray}{l}
D^{mn}=\left\{\begin{matrix} \eta^{mn} \quad&m,n\in\{0,1,\ldots,p\} \\ 
-\eta^{mn}\quad&m,n\in\{p+1,\ldots,9\} \\
0 \quad& {\rm otherwise} \end{matrix}\right. \ ,
\label{eq::boundary_matrix}
\end{IEEEeqnarray} 
which fulfils
\begin{IEEEeqnarray}{rCl}
D^{-1}=D^T=D\ .
\end{IEEEeqnarray}
As described above, only the momentum parallel to the brane is conserved. So for momentum conservation, we have
\begin{IEEEeqnarray}{l}
\sum_{i=1}^N (k_i+D{\cdot}k_i)^m=0\ .
\end{IEEEeqnarray}
Concerning the matrices $M^{\alpha}_{\hphantom \alpha \beta}$ and $N_\alpha^{\hphantom \alpha \beta}$ we will need that
\be
N_{\alpha}^{\hphantom \alpha \gamma} M^\beta_{\hphantom \beta \gamma} = \delta_\alpha^{\hphantom \alpha \beta} \ & {\rm i.e.} & \ N = (M^T)^{-1} \ , \label{NMdelta} \\
M^\gamma_{\hphantom \gamma \alpha} \gamma^m_{\gamma \delta} M^\delta_{\hphantom \delta \beta} = D^m_{\hphantom m n} \gamma^n_{\alpha \beta} = N_\alpha^{\hphantom \alpha \gamma} \gamma^m_{\gamma \delta} N_\beta^{\hphantom \beta \delta}\ & {\rm i.e.} & \ M^T \gamma^m M =  D^m_{\hphantom m n} \gamma^n = N \gamma^m N^T ,  \label{eq::gammadown} \\
M^\alpha_{\hphantom \alpha \gamma} \gamma^{m \gamma \delta} M^\beta_{\hphantom \beta \delta} = D^m_{\hphantom m n} \gamma^{n \alpha \beta} = N_\gamma^{\hphantom \gamma \alpha} \gamma^{m \gamma \delta} N_\delta^{\hphantom \delta \beta} \ & {\rm i.e.} & \ M \widehat \gamma^m M^T =  D^m_{\hphantom m n} \widehat \gamma^n = N^T \widehat \gamma^m N .\label{eq::gammaup} 
\ee
On the right hand side we give the corresponding relations in matrix notation. In order to indicate whether we are talking about the gamma matrices with lower or upper indices, we use the symbols $\gamma^m$ and $\widehat \gamma^m$, respectively. This is done in this subsection in order to make it easier to follow the ensuing discussion of the right-moving vertex operators. In the rest of the article, the position of the indices can be inferred from the context if they are not given explicitly. The relations \eqref{NMdelta}-\eqref{eq::gammaup} are derived in appendix \ref{app:MN}. A similar analysis in the context of the RNS formalism can be found for instance in appendix B of \cite{Garousi:1996ad}. 

We can make the replacements \eqref{eq::replace} in the right-moving parts of the vertex operators \eqref{closed_vops}. In this way the right-moving superfields can be expressed in terms of $X$ and $\theta$. We would like to demonstrate this using $\overline A_{\alpha}[\overline\xi,k](\overline X,\overline \theta)$ as an example. This can be rewritten according to 
\begin{IEEEeqnarray}{rCl}
	\overline A_{\alpha}[\overline\xi,k](\overline X,\overline \theta)&=&\overline A_{\alpha}[\overline\xi,k](D{\cdot}X,M \theta)\nonumber\\
	&=&e^{ik{\cdot}D{\cdot}X}\left\{\overline\xi_m(\gamma^m M\theta)_\alpha-\frac{1}{16}(\gamma_{p} M \theta)_\alpha(\theta M^{T} \gamma^{mnp} M\theta) ik_{[m}\overline\xi_{n]}\right\}\IEEEeqnarraynumspace\nonumber\\
	&=&e^{ik{\cdot}D{\cdot}X}\biggl\{\overline\xi_m(\underbrace{(M^T)^{-1} M^{T}}_{=\mathds{1}}\gamma^m M \theta)_\alpha 
%\nonumber\\
%	&&
	-\frac{1}{16}(\underbrace{(M^T)^{-1} M^{T}}_{=\mathds{1}}\gamma_{p} M\theta)_\alpha(\theta M^{T}\gamma^{mnp} M\theta) ik_{[m}\overline\xi_{n]}\biggl\}\nonumber\\
	&=&e^{ik{\cdot}D{\cdot}X} ((M^T)^{-1})^{\hphantom\alpha\beta}_{\alpha} \biggl\{(D{\cdot}\overline\xi )_m(\gamma^m \theta)_\beta - \frac{1}{16}(\gamma_{p} \theta)_\beta(\theta\gamma^{mnp} \theta)i(D{\cdot}k)_{[m}(D{\cdot}\overline\xi)_{n]}\biggl\}\nonumber\\
	&=&((M^T)^{-1})^{\hphantom\alpha\beta}_{\alpha}A_{\beta}[D{\cdot}\overline\xi,D{\cdot}k](X,\theta)\ ,\label{eq::replace1}
\end{IEEEeqnarray}
where we used 
\be
	M^{T}\gamma^{mnp} M = M^T \gamma^{[m} \widehat \gamma^{n} \gamma^{p]} M & = & M^T \gamma^{[m} M M^{-1} \widehat \gamma^{n} (M^T)^{-1} M^T \gamma^{p]} M \nonumber \\
	& = & M^T \gamma^{[m} M N^T \widehat \gamma^{n} N M^T \gamma^{p]} M  \nonumber \\ 
	& = & D^{[m}_{\hphantom m q} D^n_{\hphantom n r} D^{p]}_{\hphantom p s} \gamma^q \widehat \gamma^r \gamma^s = D^{m}_{\hphantom m q} D^n_{\hphantom n r} D^{p}_{\hphantom p s} \gamma^{[q} \widehat \gamma^r \gamma^{s]}  \nonumber \\
	& = & D^{m}_{\hphantom m q} D^n_{\hphantom n r} D^{p}_{\hphantom p s} \gamma^{qrs}\ .
\ee
Moreover, note that in a contraction of fermions like $(\overline \theta \gamma^{mnp} \overline \theta)$ or $\overline\lambda^{\alpha}\overline A_{\alpha}$ etc., the left spinor is implicitly a transposed spinor. This explains the appearance of $M^T$ in the second row of \eqref{eq::replace1}. For the other superfields we find analogously
\begin{IEEEeqnarray}{rCl}
\begin{IEEEeqnarraybox}[][c]{rCl}
\IEEEstrut
\overline A_m[\overline\xi,k](\overline X,\overline\theta)&=&D_{m}^{\hphantom mn}A_n[D{\cdot}\overline\xi,D{\cdot}k](X,\theta)\ ,\\
\overline W^\alpha[\overline\xi,k](\overline X,\overline\theta) &=& M_{\hphantom\alpha\beta}^{\alpha} W^\beta[D{\cdot}\overline\xi,D{\cdot}k](X,\theta)\ ,\\
\mathcal{\overline F}_{mn}[\overline\xi,k](\overline X,\overline\theta)&=&D_{m}^{\hphantom m p}D_{n}^{\hphantom n q} \mathcal{F}_{pq}[D{\cdot}\overline\xi,D{\cdot}k](X,\theta)\ .
\IEEEstrut
\end{IEEEeqnarraybox}\label{eq::replace2}
\end{IEEEeqnarray}
We also have to transform the remaining right-moving world-sheet fields appearing in the vertex operators. Analogously to \eqref{eq::replace}, we have (cf.\ also appendix \ref{app:MN}) 
\begin{IEEEeqnarray}l
\begin{IEEEeqnarraybox}[][c]{rClCrClCrClCrCl}
\IEEEstrut
%\overline X^m&=&D^m_{\hphantom mn}X^n,&\qquad&\overline\theta^\alpha&=&M^\alpha_{\hphantom\alpha\beta}\theta^\beta,&\qquad&
\overline d_\alpha & = & ((M^T)^{-1})^{\hphantom\alpha\beta}_\alpha d_\beta\ , & \quad & \overline \lambda^\alpha & = & M^\alpha_{\hphantom\alpha\beta}\lambda^\beta\ , & \quad & \overline \Pi^m & = & D^{m}_{\hphantom mn} \Pi^n\ , & \quad & \overline N^{mn} & = & D^{m}_{\hphantom m p}D^{n}_{\hphantom n q} N^{pq}\ .
\IEEEstrut
\end{IEEEeqnarraybox}\label{eq::replace3}
\end{IEEEeqnarray}
Putting everything together we obtain the right-moving part of the vertex operators as
\begin{IEEEeqnarray}{rCl}
\overline V^{(0)}(\overline z)&=&\Big( \overline\lambda^{\alpha}\overline A_{\alpha}[\overline\xi,k](\overline X,\overline\theta)\Big)(\overline z) = \Big( \lambda^\alpha A_\alpha[D{\cdot}\overline\xi,D{\cdot}k](X,\theta)\Big)(\overline z)\ ,
\label{eq::redefinedvertex1}\\
\overline V^{(1)}(\overline z)&=& \Big( \overline\partial\overline\theta^\alpha \overline A_\alpha[\overline\xi,k](\overline X,\overline\theta)+\overline\Pi^m\overline A_m[\overline\xi,k](\overline X,\overline \theta)\nonumber\\
&&+\overline d_\alpha \overline W^\alpha[\overline\xi,k](\overline X,\overline\theta)+\frac12\overline N^{mn}\mathcal{\overline F}_{mn}[\overline\xi,k](\overline X,\overline\theta) \Big)(\overline z) \nonumber\\
&=&\Big( \overline\partial\theta^\alpha A_\alpha[D{\cdot}\overline\xi,D{\cdot}k](X,\theta)+\Pi^mA_m[D{\cdot}\overline\xi,D{\cdot}k](X,\theta)\nonumber\\
&& + d_\alpha W^\alpha[D{\cdot}\overline\xi,D{\cdot}k](X,\theta)+\frac12N^{mn}{\mathcal F}_{mn}[D{\cdot}\overline\xi,D{\cdot}k](X,\theta)\Big)(\overline z)\ .
\label{eq::redefinedvertex2}
\end{IEEEeqnarray}

%%%%%%%%%%%%%%%%%%%%%%%%%%%%%%%%%%%%%%

\subsection{Calculating correlators}

The prescription to calculate closed string amplitudes on the sphere and open string amplitudes on the disk is well known and tested in the pure spinor formalism. Both world-sheets do not have any moduli and their conformal Killing vectors (CKVs) allow the fixing of three closed or open vertex operators, respectively. For an $n$-point function with $n \geq 3$, it is a convenient choice to fix the vertex operators $i=1,n-1$ and $n$ to the positions $z_1, z_{n-1}$ and $z_n$:
\begin{IEEEeqnarray}{l}
\begin{IEEEeqnarraybox}[][c]{rCl}
\mathcal A^\text{closed}_{S^2}(1,2,\ldots,n)&=& \left\langle V^{(0,0)}_1(z_1,\overline z_1) \prod_{i=2}^{n-2}\int\mathrm d^2z_i V^{(1,1)}_i(z_i,\overline z_i) V^{(0,0)}_{n-1}(z_{n-1},\overline z_{n-1})V^{(0,0)}_{n}(z_n,\overline z_n)\right\rangle , \\
\mathcal A^\text{open}_{D_2}(1,2,\ldots,n)&=& \left\langle V^{(0)}_1(z_1)  \prod_{i=2}^{n-2}\int_{z_{i-1}}^{z_{n-1}}\mathrm dz_i V^{(1)}_i(z_i) V^{(0)}_{n-1}(z_{n-1})V^{(0)}_{n}(z_n)\right\rangle . \IEEEeqnarraynumspace
\end{IEEEeqnarraybox}
\label{eq::treepres}
\end{IEEEeqnarray}
For open superstrings the integrated vertex operator positions are ordered, such that $z_1\le z_2\le z_3\le\ldots\le z_{n-2}\leq z_{n-1}\leq z_n$, and integrated over the corresponding parts of the real axis (if we perform the calculation on the upper half plane). Other orderings of the vertex operators can then be inferred via relabelling. Almost all tree level calculations performed in the pure spinor formalism so far are of these two types. The only exceptions that we are aware of can be found in \cite{Alencar:2008fy, Alencar:2011tk} which calculate 3-point functions on the disk with one closed and two open string vertex operators. However, as we mentioned in the introduction, the case with only closed string vertex operators on the disk is more subtle, given that fixing the CKG of the disk only allows to fix one and a half vertex operators. Moreover, the disk amplitude with a single closed string vertex operator would naively vanish with the usual tree level prescription as it does not contain three factors of $\lambda^\alpha$ (cf.\ section \ref{sec::zeromode} below). Both of these issues will be addressed in the following. 

%%%%%%%%%%%%%%%%%%%%%%%%%%%%%%%%%%%%%%

\subsubsection{Integrating out the non-zero modes}\label{sec::non-zero}

In this section we follow relatively closely the presentation in \cite{Schlotterer:2011psa}. We introduce the shorthand notation $A_\alpha(z)=A_\alpha\left(X(z),\theta(z)\right)$ and similarly for the other superfields. Moreover, we denote by $\mathcal V$ any of the superfields, i.e.\ $\mathcal V\in\left\{A_\alpha, A_m, W^\alpha, \mathcal F_{mn}\right\}$. 

The $h=1$ primaries $\partial\theta^\alpha,\Pi^m,d_\alpha,N^{mn}$ appearing in the integrated vertex operator \eqref{V1} do not have any zero modes at tree level. Thus, they can be integrated out by applying Wick's theorem, employing the OPEs
\begin{IEEEeqnarray}{rCl}
\Pi^m(z) \mathcal V(w)&=&\frac{-ik^m\mathcal V(w)}{z-w}\ , \label{opepiv}\\
d_\alpha(z)\mathcal V(w)&=&\frac{D_\alpha\mathcal V(w)}{z-w}\ ,\\
N^{mn}(z)\lambda^\alpha A_\alpha(w)&=&-\frac12\frac{(\lambda\gamma^{mn})^\alpha}{z-w} A_\alpha(w)\ , \label{opena}
\end{IEEEeqnarray}
where in the last equality we used the fact that the matrix $\gamma^{mn}$ is antisymmetric in the spinor indices: $(\gamma^{mn})^{\hphantom\alpha\beta}_{\alpha}=-(\gamma^{mn})^\beta_{\hphantom{\beta}\alpha}$. It turns out that $\partial\theta^\alpha$ does not contribute to the closed string two- and one-point functions that we are going to calculate, given that it only has a singular OPE with $d_\alpha$. In the two-point function there is only one integrated vertex operator and for the one-point function its potential contribution vanishes due to the zero mode integration, cf.\ section \ref{sec::zeromode}. Using \eqref{opepiv}-\eqref{opena} and the superfield equations of motion \eqref{eq::eoms_superfields}, one infers the OPE of $V^{(0)}_i(z_i)$ with $V^{(1)}_j(z_j)$ via
\begin{IEEEeqnarray}{rCl}
V^{(0)}_i(z_i)\Pi^mA_m^j(z_j)&=&\frac{1}{z_{ji}}(-ik^m_i)\lambda^\alpha A^i_\alpha A^j_m(z_i)=-\frac{1}{z_{ji}}ik_i{\cdot}A_jV^{(0)}_i(z_i)\ ,\label{eq::build_1}\\
V^{(0)}_i(z_i)d_\alpha W^\alpha_j(z_j)&=&-\frac{1}{z_{ji}}\lambda^\alpha \left(D_\beta A^i_\alpha\right) W^\beta_j(z_i) \nonumber \\
&=&-\frac{1}{z_{ji}}\lambda^\alpha(-D_\alpha A^i_\beta+\gamma^m_{\alpha\beta}A^i_m)W^\beta_j(z_i) \nonumber \\
&=&\frac{1}{z_{ji}}\left[(Q A^i_\alpha) W^\alpha_j-A^i_m(\lambda\gamma^mW_j)\right](z_i)\ ,\label{eq::build_2}\\
V^{(0)}_i(z_i)\frac12N^{mn}\mathcal F^j_{mn}(z_j)&=&-\frac{1}{4z_{ji}}\lambda^\alpha (\gamma^{mn})^{\hphantom{\alpha}\beta}_{\alpha}A^i_\beta\mathcal F^j_{mn}(z_i)=-\frac{1}{z_{ji}}A^i_\alpha(QW^\alpha_j)(z_i)\ . \label{eq::build_3}
\end{IEEEeqnarray}
The two terms $(QA^i_\alpha W^\alpha_j)$ and $-A^i_\alpha(QW_j^\alpha)$ can be combined to the BRST exact expression $Q(A_iW_j)$, which does not contribute to the two-point function that we are going to calculate. Hence, we can effectively use
\begin{IEEEeqnarray}{rCl}
V^{(0)}_i(z_i)V^{(1)}_j(z_j)&=&\frac{1}{z_{ji}}\left[-ik_i\cdot A_j V^{(0)}_i-A^i_m(\lambda\gamma^mW_j)\right](z_i)\ .
\label{OPEV0V1}
\end{IEEEeqnarray}
This is all we need for the two-point function, given that there is only one integrated vertex operator and the non-zero modes of the unintegrated vertex operators only interact via their plane wave-factors. Similarly, apart from the plane-wave factors, the terms on the right hand side of \eqref{OPEV0V1} only depend on $\lambda$ and $\theta$, which do not have any non-trivial OPEs with the remaining unintegrated vertex operators in the two-point calculation. Thus, aside from the plane wave factors, only the zero modes of $\lambda$ and $\theta$ will contribute in the remaining computation of the two-point function after employing \eqref{OPEV0V1}. The zero mode integration is discussed in the next section. Here, we would just like to recall that the interaction of the plane waves contributes the Koba-Nielsen factor, which at tree level takes the form
\begin{IEEEeqnarray}{rCl}
\mathcal I_n=\left\langle\prod_{i=1}^{n}e^{ik_i{\cdot}X(z_i)}\right\rangle = C^X_{D_2} \prod_{\substack{i,j=1\\i \neq j}}^{n}\left( z_i-z_j \right)^{k_i{\cdot}k_j / 2}\ .
\end{IEEEeqnarray}
The constant $C^X_{D_2}$ arises from the path integral over the non-zero modes of $X^m$, cf.\ \cite{Polchinski:1998rq}.\footnote{Strictly speaking, our usage of the symbol $\langle \ldots \rangle$ is a bit ambiguous. Most of the time, we use it like here to denote a path integral expectation value (and, thus, it contains constants like this $C^X_{D_2}$ from the path integral over the world-sheet fields). However, sometimes (i.e.\ if $\langle \ldots \rangle$ only has two world-sheet fields as arguments) we also use it to denote the Green's function, cf.\ for instance \eqref{eq::correlator} and \eqref{eq::correlator2ptdisk}.}

%%%%%%%%%%%%%%%%%%%%%%%%%%%%%%%%%%%%%%

\subsubsection{Integrating out the zero modes}
\label{sec::zeromode}

At tree level only the conformal weight zero fields have zero modes. These are $X^m,\theta^\alpha$ and $\lambda^\alpha$. The evaluation of the $X^m$ zero modes is the same as for the RNS superstring and gives a momentum preserving $\delta$-function. The evaluation of the correlator of the remaining zero modes of $\theta^\alpha$ and $\lambda^\alpha$ is more subtle. 

There exist actually two different zero mode prescriptions for tree level calculations in the literature which are usually equivalent but can differ for very low point functions, as we will see below. The usual prescription fixes the $PSL(2,\mathbb{C})$ invariance on the sphere and the $PSL(2,\mathbb{R})$ invariance on the disk by utilizing three unintegrated vertex operators, $V^{(0)}_i \overline V^{(0)}_i=\lambda^\alpha A^i_\alpha \overline \lambda\, \! ^{\beta} \overline A\, \! ^i_{\beta}$ for closed strings on the sphere or $V^{(0)}_i=\lambda^\alpha A^i_\alpha$ for open strings on the disk. When dealing with closed strings on the disk, one would also use $V^{(0)}_i=\lambda^\alpha A^i_\alpha$, albeit for three factors of the closed string vertex operators \eqref{closed_vops} (after rewriting the right-movers according to \eqref{eq::redefinedvertex1}). This is the case we are going to discuss and apply in the following. 

After performing the integration of the non-zero modes of $\partial\theta^\alpha, \Pi^m, d_\alpha$ and $N^{mn}$ as described in the last subsection, one is left with an expression cubic in the ghosts $\lambda^\alpha$:
\begin{IEEEeqnarray}{l}
\left\langle V^{(0)}_1(z_1) \prod_{i=2}^{n-2} V^{(1)}_i(z_i)V^{(0)}_{n-1}(z_{n-1})V^{(0)}_n(z_n)\right\rangle=\left\langle\lambda^\alpha\lambda^\beta\lambda^\gamma f_{\alpha\beta\gamma}(\theta;z_i)\right\rangle_0 . 
\label{eq::zero_mode_expression}
\end{IEEEeqnarray}
The subscript on $\langle \ldots \rangle_0$ indicates that the bracket on the right hand side denotes a zero mode prescription, as all the non-zero modes are already integrated out. The argument of $\langle \ldots \rangle_0$ in \eqref{eq::zero_mode_expression} has a finite power series expansion in $\theta^\alpha$ and it was argued in \cite{Berkovits:2000fe} that only terms involving five powers of $\theta$ contribute (at ghost number 3). Given that the tensor product of three $\lambda$ and five $\theta$ contains a unique scalar, all terms of this type are proportional to each other (cf.\ the nice discussion in appendix A of \cite{Mafra:2009wq}) and are determined by\footnote{We chose the normalization for convenience in order to simplify the explicit factors arising in the two-point calculations of sec.\ \ref{2pt}. A different choice of this normalization could be absorbed in the constant $C_{D_2}$ appearing in the final result, cf.\ \eqref{final_2pt} below. We follow here the convention used for instance in \cite{Mafra:2014gja}.}
\begin{IEEEeqnarray}{rCl}
\langle(\lambda\gamma^m\theta)(\lambda\gamma^n\theta)(\lambda\gamma^p\theta)(\theta\gamma_{mnp}\theta)\rangle_0 = 2880\ .\label{eq::zmp}
\end{IEEEeqnarray}
Two further examples of zero mode expressions, which follow from \eqref{eq::zmp} and which are needed for the two-point function calculation of section \ref{2pt}, are
\begin{IEEEeqnarray}{rCl}
\left\langle(\lambda\gamma^m\theta)(\lambda\gamma^n\theta)(\lambda\gamma^p\theta)(\theta\gamma_{rst}\theta)\right\rangle_0 &=&24\delta_{r}^{[m}\delta^{n}_{s}\delta^{p]}_{t}\ , \label{eq::zmp1}\\
\left\langle(\lambda\gamma_m\theta)(\lambda\gamma_s\theta)(\lambda\gamma^{ptu}\theta)(\theta\gamma_{fgh}\theta)\right\rangle_0 &=&\frac{288}{7}\delta^{[p}_{[m}\eta_{s][f}\delta^{t}_{g}\delta^{u]}_{h]}\ .
\label{eq::zmp2}
\end{IEEEeqnarray}
These and additional zero mode expressions can be found, for instance, in appendix A of \cite{Mafra:2009wq}.

As discussed in \cite{Berkovits:2016xnb} there is an alternative zero mode prescription at tree level (cf.\ also \cite{Lee:2006pa}), i.e. 
\begin{IEEEeqnarray}{rCl}
\langle\mathds{1}\rangle_0 = 1\ .
\label{eq::new_zmp}
\end{IEEEeqnarray}
In this zero mode prescription only terms involving no $\lambda$- and $\theta$-zero modes contribute. In other words, while at ghost number three, the unique element in the BRST cohomology is proportional to $\theta^5$, at ghost number zero, it is proportional to the unit operator. 

Using this alternative zero mode prescription, only integrated vertex operators contribute to scattering amplitudes. Further, using \eqref{eq::new_zmp} we can simplify the integrated vertex operator if we are looking at a purely bosonic scattering amplitude. In that case the only contribution to the amplitude comes from \cite{Berkovits:2016xnb}
\begin{IEEEeqnarray}{rCL}
V^{(1)}(z)=\partial X^m A_m(z) - \frac12M^{mn}\mathcal F_{mn}(z)\ ,
\label{V1Mmn}
\end{IEEEeqnarray}
where $M^ {mn}$ is the Lorentz current \eqref{eq::lorentz_current} of the pure spinor formalism.\footnote{Compared to \cite{Berkovits:2016xnb} we adjusted the sign of the $M^{mn}$-term to our conventions.}

We will need this alternative zero-mode prescription for calculating the closed string one-point function on the disk in section \ref{1pt}.

%%%%%%%%%%%%%%%%%%%%%%%%%%%%%%%%%%%%%%%%%%%

\section{Closed string two-point function on the disk}
\label{2pt}

We now want to calculate the elastic scattering of two massless bosonic states off a Dirichlet $p$-brane in a flat background using the pure spinor formalism. More precisely, we consider states which would arise from the NSNS sector in the RNS formalism (i.e.\ the graviton, dilaton and Kalb-Ramond field). We calculate the amplitude by inserting two closed string vertex operators with appropriate boundary conditions on the disk. The corresponding calculation in the RNS formalism can be found in \cite{Garousi:1996ad,Hashimoto:1996bf}. Further we recommend \cite{Aldi:2020dvw}, where a rather detailed presentation of the RNS calculation is given.

The two closed strings have polarizations $\epsilon_1$ and $\epsilon_2$ and momenta $k_1$ and $k_2$. These strings are massless and have transverse polarizations, i.e.
\begin{IEEEeqnarray}l
k_j^2=0\ , \qquad \epsilon_{jmn}k_j^m=\epsilon_{jmn}k_j^n=0\ .
\end{IEEEeqnarray}

As a first step to calculate the scattering of two massless closed strings on the disk in the pure spinor formalism, we need to formulate a correlation function. For closed strings on the disk, we can not just use the tree level prescription \eqref{eq::treepres}. Whereas on the sphere the six conformal Killing vectors allow us to fix three vertex operator positions, the situation on the disk is different. Because we have only three conformal Killing vectors, we can only fix one and a half vertex operator positions. Since we have two vertex operator insertions at points $z_1=x_1+iy_1$ and $z_2=x_2+iy_2$, we are going to fix the positions $x_1=0,x_2=0$ and $y_2=1$ and keep the integration over $y_1$, which we will call $y$ in the following. 

The form of the correlator in this case can be inferred by employing the methods of \cite{Hoogeveen:2007tu,Hoogeveen:2010yfa} (which are based on \cite{Craps:2005wk}). As the discussion in those papers is rather technical and the final result in \eqref{eq::2ptdisk_2} below is rather plausible (and similar to the RNS string, cf.\ formula (2.12) in \cite{Aldi:2020dvw}), we refrain from reviewing the details of \cite{Hoogeveen:2007tu,Hoogeveen:2010yfa,Craps:2005wk}. Suffice it to say that \cite{Hoogeveen:2007tu} introduces anti-commuting BRST partners $\zeta_j = \zeta_j^x + i \zeta_j^y$ of the complex vertex operator positions $z_j$ (where $j$ enumerates the vertex operators) and uses them to write the vertex operators as
\begin{IEEEeqnarray}{rCL}
V_j(z_j, \zeta_j) = V^{(0)}_j(z_j) + \zeta_j V^{(1)}_j(z_j) \ .
\end{IEEEeqnarray}
Applying the formulas of \cite{Hoogeveen:2007tu} to the case at hand leads then to\footnote{To be a bit more concrete, formula (5.8) in \cite{Hoogeveen:2007tu} would take the form $L_2=\pi^1_xx_1+\pi^2_xx_2+\pi^2_y(y_2-1)-b^1_xc^x(x_1)-b^2_xc^x(x_2)-b^2_y c^y (y_2)-p^1_x\zeta^x_1-p^2_x\zeta^x_2-p^2_y\zeta^y_2+\beta^1_x\gamma^x(x_1)+\beta^2_x\gamma^x(x_2)+\beta^2_y\gamma^y(y_2)$, where $\pi^j_a$ and $p^j_a$ are auxiliary fields, which lead to the delta functions in the first line of \eqref{eq::2ptdisk_2} when integrated out, while $b^j_a$ and $\beta^j_a$ are antighosts which lead to factors of $c^a$ and $\delta (\gamma^a)$ when integrated out. Here, $c^a$ and $\gamma^a$ are the diffeomorphism ghosts and their (commuting) BRST partners. However, as shown in appendix B of \cite{Hoogeveen:2007tu} all the ghost contributions cancel in the final correlator, thus leaving \eqref{eq::2ptdisk_2} when the dust settles.}
\begin{IEEEeqnarray*}{rCl}
\IEEEeqnarraymulticol{3}{l}{\mathcal{A}^\text{closed}_{D_2}(1,2)=} \nonumber \\
&=&\frac{g_c^2\tau_p}{2}\prod_{j=1}^2\int\mathrm dx_j\,\mathrm dy_j\int\mathrm d\zeta^x_j\,\mathrm d\zeta^y_j\,\delta(x_1)\delta(x_2)\delta(y_2-1)\delta(\zeta^x_1)\delta(\zeta^x_2)\delta(\zeta_2^y)\IEEEyesnumber\label{eq::2ptdisk_1}  \nonumber \\
&& \hspace{4cm} \times\left\langle V_1(z_1,\zeta_1) \overline V_1(\overline z_1,\overline\zeta_1) V_2(z_2,\zeta_2)\overline V_2(\overline z_2,\overline\zeta_2)\right\rangle\\
&=&\frac{g_c^2\tau_p}{2}\prod_{j=1}^2\int\mathrm dx_j\,\mathrm dy_j\int\mathrm d\zeta^x_j\,\mathrm d\zeta^y_j\,\delta(x_1)\delta(x_2)\delta(y_2-1)\delta(\zeta^x_1)\delta(\zeta^x_2)\delta(\zeta_2^y)\\
&&\hspace{0.5cm} \times\biggl\langle\left(V^{(0)}_1(x_1+iy_1)+(\zeta^x_1+i\zeta^y_1)V^{(1)}_1(x_1+iy_1)\right)
% \\ && \times
\left(\overline V^{(0)}_1(x_1-iy_1)+(\zeta^x_1-i\zeta^y_1)\overline V^{(1)}_1(x_1-iy_1)\right)\\
&& \hspace{0.5cm} \times\left(V^{(0)}_2(x_2+iy_2)+(\zeta^x_2+i\zeta^y_2)V^{(1)}_2(x_2+iy_2)\right)
% \\ && \times
\left(\overline V^{(0)}_2(x_2-iy_2)+(\zeta^x_2-i\zeta^y_2)\overline V^{(1)}_2(x_2-iy_2)\right)\biggl\rangle\\
&=&\frac{g_c^2\tau_p}{2} \int\mathrm dy_1\int\mathrm d\zeta^y_1\,\biggl\langle\left(V^{(0)}_1(iy_1)+i\zeta^y_1V^{(1)}_1(iy_1)\right)\left(\overline V^{(0)}_1(-iy_1)-i\zeta^y_1\overline V^{(1)}_1(-iy_1)\right)
%\IEEEeqnarraynumspace \\ &&\times 
V^{(0)}_2(i)\overline V^{(0)}_2(-i)\biggl\rangle\IEEEyesnumber\label{eq::zeta} \nonumber \\
&=&\frac{g_c^2\tau_p}{2}\int\mathrm dy\,\biggl\langle i\left(V^{(0)}_1(iy)\overline V^{(1)}_1(-iy)+V_1^{(1)}(iy)\overline V^{(0)}_{1}(-iy)\right)V^{(0)}_2(i)\overline V^{(0)}_2(-i)\biggl\rangle\ .\IEEEyesnumber\IEEEeqnarraynumspace
\label{eq::2ptdisk_2}
\end{IEEEeqnarray*}
In the last equality we performed the Grassmann integral $\int d\zeta^y_1\, \zeta^y_1 = 1$ and used the fact that $V^{(0)}_i$ and $\zeta^y_1$ are anticommuting. Moreover, we inserted the coupling constant $g_c = e^{\langle \Phi \rangle}$ for closed strings and the physical tension $\tau_p = e^{-\langle \Phi \rangle} (2 \pi)^{-p} \alpha'^{-(p+1)/2}$ of the D$p$-brane in the definition of the amplitude.

The gauge fixing in \eqref{eq::2ptdisk_2} is not complete yet. The amplitude is still invariant under the $PSL(2,\mathbb R)$ transformation ${z_1}\rightarrow-\frac{1}{z_1}$ which acts on $y$ according to $y \rightarrow \frac{1}{y}$ (to account for this discrete unfixed conformal Killing transformation, we inserted the factor $1/2$ in \eqref{eq::2ptdisk_2}). This transformation maps the interval $0<y<1$ to $1<y<\infty$. We can fix also this discrete symmetry by restricting to only one of the two intervals and multiplying the result by $2$. For concreteness we choose $0<y<1$, i.e.\
\begin{IEEEeqnarray}{l}
\mathcal{A}^\text{closed}_{D_2}(1,2)=\nonumber\\
= i g_c^2\tau_p\int_0^1\mathrm dy\,\biggl\langle \left(V^{(0)}_1(iy)\overline V^{(1)}_1(-iy)+V_1^{(1)}(iy)\overline V^{(0)}_{1}(-iy)\right)V^{(0)}_2(i)\overline V^{(0)}_2(-i)\biggl\rangle\ . 
\label{eq::2ptdisk}
\end{IEEEeqnarray}

The two summands in \eqref{eq::2ptdisk} are actually equal, as we show in appendix \ref{app:independent}, and, thus, we finally obtain
\begin{IEEEeqnarray}{l}
\mathcal{A}^\text{closed}_{D_2}(1,2) = 2 i g_c^2\tau_p\int_0^1\mathrm dy\,\biggl\langle V^{(0)}_1(iy) \overline V^{(1)}_1(-iy) V^{(0)}_2(i)\overline V^{(0)}_2(-i)\biggl\rangle\ . 
\label{eq::2ptdisk_simple}
\end{IEEEeqnarray}

Using \eqref{eq::redefinedvertex1} and \eqref{eq::redefinedvertex2}, the amplitude \eqref{eq::2ptdisk_simple} becomes a standard correlation function in the pure spinor formalism (i.e.\ one which can be evaluated using the standard correlators \eqref{eq::correlator})
\begin{IEEEeqnarray}{rCl}
\IEEEeqnarraymulticol{3}{l}{\mathcal{A}^\text{closed}_{D_2}(1,2)=} \nonumber \\
&=& 2 i g_c^2\tau_p \int_0^1\mathrm dy\,\biggl\langle (\lambda A_1[\xi_1,k_1])(iy)\biggl(\overline\partial\theta^\alpha A_{1\alpha}[D{\cdot}\overline\xi_1,D{\cdot}k_1]+\Pi^mA_{1m}[D{\cdot}\overline\xi_1,D{\cdot}k_1] 
\label{eq::redefined_amplitude} \IEEEyesnumber\\
&&+d_\alpha W^{\alpha}_1[D{\cdot}\overline\xi_1, D{\cdot}k_1]+\frac12N^{mn} {\mathcal F}_{1mn}[D{\cdot}\overline\xi_1,D{\cdot}k_1]\biggl)(-iy) (\lambda A_2[\xi_2,k_2])(i)(\lambda A_2[D{\cdot}\overline\xi_2,D{\cdot}k_2])(-i)\biggl\rangle\ . \IEEEeqnarraynumspace \nonumber 
\end{IEEEeqnarray}
The expression in \eqref{eq::redefined_amplitude} can be evaluated following the steps in section \ref{PSF}, resulting in
\begin{IEEEeqnarray}{rCl}
\IEEEeqnarraymulticol{3}{l}{\mathcal{A}^\text{closed}_{D_2}(1,2)= 2 g_c^2\tau_pC_{D_2}\int_0^1\mathrm dy\, |2 i y|^{k_1{\cdot}D{\cdot}k_1} |2 i|^{k_2{\cdot}D{\cdot}k_2} |i-iy|^{2k_1{\cdot}k_2}|i+iy|^{2k_1{\cdot}D{\cdot}k_2}}\nonumber\\
&&\times\biggl\langle-\frac{1}{2y}\Big(\underbrace{-i(\lambda A_{1}[\xi_1,k_1])k_{1}{\cdot}A_{1}[D{\cdot}\overline\xi_1,D{\cdot}k_1](\lambda A_{2}[\xi_2,k_2])(\lambda A_{2}[D{\cdot}\overline\xi_2,D{\cdot}k_2])}_{=d'_1}\nonumber\\
&&\underbrace{- A_{1m}[\xi_1,k_1](\lambda\gamma^mW_1[D{\cdot}\overline\xi_1,D{\cdot}k_1])(\lambda A_{2}[\xi_2,k_2])(\lambda A_{2}[D{\cdot}\overline\xi_2,D{\cdot}k_2])}_{=d''_1}\Big)\nonumber\\
&&-\frac{1}{1+y}\Big(\underbrace{-i(\lambda A_{1}[\xi_1,k_1])k_{2}{\cdot}A_{1}[D{\cdot}\overline\xi_1,D{\cdot}k_1](\lambda A_{2}[\xi_2,k_2])(\lambda A_{2}[D{\cdot}\overline\xi_2,D{\cdot}k_2])}_{=d'_2}\nonumber\\
&&\underbrace{-(\lambda A_1[\xi_1,k_1])(\lambda\gamma^mW_1[D{\cdot}\overline\xi_1,D{\cdot}k_1])A_{2m}[\xi_2,k_2](\lambda A_{2}[D{\cdot}\overline\xi_2,D{\cdot}k_2])}_{=d''_2}\Big)\nonumber\\
&&+\frac{1}{1-y}\Big(\underbrace{-i(\lambda A_{1}[\xi_1,k_1])k_{2}{\cdot}D{\cdot}A_{1}[D{\cdot}\overline\xi_1,D{\cdot}k_1](\lambda A_{2}[\xi_2,k_2])(\lambda A_{2}[D{\cdot}\overline\xi_2,D{\cdot}k_2])}_{=d'_3}\IEEEeqnarraynumspace\nonumber\\
&&\underbrace{+(\lambda A_{1}[\xi_1,k_1])(\lambda\gamma^m W_1[D{\cdot}\overline\xi_1,D{\cdot}k_1])(\lambda A_2[\xi_2,k_2])A_{2m}[D{\cdot}\overline\xi_2,D{\cdot}k_2]}_{=d''_3}\Big) \biggl\rangle_0
\label{eq::intermediateresult}\\
&=&2 g_c^2\tau_pC_{D_2}\int_0^1\mathrm dy\,\left(\frac{4y}{(1+y)^2}\right)^{k_1{\cdot}D{\cdot}k_1}\left(\frac{(1-y)^2}{(1+y)^2}\right)^{k_1{\cdot}k_2} \left(- \frac{d_1}{2y} - \frac{d_2}{1+y} + \frac{d_3}{1-y} \right)\ ,
\label{eq::intermediateresult0.1}
\end{IEEEeqnarray}
where in the last equality we used $k_1{\cdot}D{\cdot}k_1 = k_2{\cdot}D{\cdot}k_2$ and we defined $d_i=d'_i+d''_i$. The $d_i$ are kinematic factors, which are calculated by using the first zero mode prescription described in section \ref{sec::zeromode} and fulfil the identity $d_1+d_2+d_3=0$, which can be shown by explicitly computing the $d_i$.\footnote{We used Cadabra2 \cite{Cadabra} to perform these calculations.} Note that we dropped a factor $(2\pi)^{p+1}\delta^{p+1}\! \left(k_1+D{\cdot}k_1+k_2+D{\cdot}k_2\right)$ in equation \eqref{eq::intermediateresult}, which comes from the $X^m$ zero-modes. This delta function describes the momentum conservation along the world-volume of the D-brane. Moreover, $C_{D_2}$ is again an overall constant arising from performing the path integral over the non-zero modes of the world-sheet fields.

It is customary to evaluate the integral in \eqref{eq::intermediateresult0.1} by performing the substitution \cite{Garousi:1996ad,Hashimoto:1996bf}
\begin{IEEEeqnarray}{rCl}
y=\frac{1-\sqrt{x}}{1+\sqrt{x}}\ ,
\end{IEEEeqnarray}
yielding
\begin{IEEEeqnarray*}{rCl}
\mathcal{A}^\text{closed}_{D_2}(1,2)&=&g_c^2\tau_pC_{D_2} \frac{\Gamma(k_1{\cdot}k_2)\Gamma(k_1{\cdot}D{\cdot}k_1)}{\Gamma(1+k_1{\cdot}k_2+k_1{\cdot}D{\cdot}k_1)}\big( d_3\, k_1{\cdot}D{\cdot}k_1 -  d_1\, k_1{\cdot}k_2 \big)\ .
\IEEEyesnumber
\end{IEEEeqnarray*}
Using momentum conservation and transversality, we can rewrite the kinematic factors. Finally, we obtain the result
\begin{IEEEeqnarray}{rCl}
\mathcal{A}^\text{closed}_{D_2}(1,2)&=&g_c^2\tau_pC_{D_2}\frac{\Gamma(-t/2)\Gamma(2q^2)}{\Gamma(1-t/2+2q^2)}\left(2q^2a_1+\frac t2a_2\right),
\label{final_2pt}
\end{IEEEeqnarray}
where $q^2=\frac12k_1{\cdot}D{\cdot}k_1$ is the momentum parallel to the world-volume of the brane and $t=-(k_1+k_2)^2=-2k_1{\cdot}k_2$ is the momentum absorbed by the $p$-brane.\footnote{Note that $d_3 \neq a_1$ and $d_1 \neq a_2$, because $d_3$ contains terms proportional to $k_1 \cdot k_2$ which contribute to the $\frac{t}{2} a_2$-term.} The kinematic factors $a_1$ and $a_2$ are defined by
\begin{IEEEeqnarray*}{rCl}
a_1&=&\text{Tr}(\epsilon_1{\cdot}D)k_1{\cdot}\epsilon_2{\cdot}k_1-k_1{\cdot}\epsilon_2{\cdot}D\epsilon_1{\cdot}k_2-k_1{\cdot}\epsilon_2{\cdot}\epsilon_1^T{\cdot}D{\cdot}k_1\\
&&-k_1{\cdot}\epsilon_2^T{\cdot}\epsilon_1{\cdot}D{\cdot}k_1-k_1{\cdot}\epsilon_2{\cdot}\epsilon_1^T{\cdot}k_2+q^2\text{Tr}(\epsilon_1{\cdot}\epsilon_2^T)+\{1\leftrightarrow 2\},\IEEEyesnumber\\
a_2&=&\text{Tr}(\epsilon_1{\cdot}D)(k_2{\cdot}D{\cdot}\epsilon_2{\cdot}D{\cdot}k_2+k_1{\cdot}\epsilon_2{\cdot}D{\cdot}k_2+k_2{\cdot}D{\cdot}\epsilon_2{\cdot}k_1)\\
&&+k_1{\cdot}D{\cdot}\epsilon_1{\cdot}D{\cdot}\epsilon_2{\cdot}D{\cdot}k_2-k_2{\cdot}D{\cdot}\epsilon_2{\cdot}\epsilon_1^T{\cdot}D{\cdot}k_1+q^2\text{Tr}(\epsilon_1{\cdot}D{\cdot}\epsilon_2{\cdot}D)\\
&&-q^2\text{Tr}(\epsilon_1{\cdot}\epsilon_2^T)-(q^2-\frac{t}{4})\text{Tr}(\epsilon_1{\cdot}D)\text{Tr}(\epsilon_2{\cdot}D)+\{1\leftrightarrow 2\}.\IEEEyesnumber
\end{IEEEeqnarray*}
In the pure spinor formalism we obtained exactly the same kinematic factors as in the RNS formalism \cite{Garousi:1996ad,Hashimoto:1996bf}, which is another indication for the equivalence of these two formalisms.

%%%%%%%%%%%%%%%%%%%%%%%%%%%%%%%%%%%%%%%%%%%

\section{Closed string one-point function on the disk}
\label{1pt}

One further interesting amplitude that was not calculated in the pure spinor formalism so far, is the closed superstring one-point function on the disk. For this amplitude, the usual zero-mode prescription, employed in the last section for the two-point function, can not be used, given that there is only a single closed string vertex operator (or alternatively two open string factors). Hence, our strategy is to make use of the alternative zero-mode prescription of \cite{Berkovits:2016xnb}, reviewed in section \ref{sec::zeromode}. More concretely, we are not going to fix the position of the single vertex operator, but divide by the volume of the CKG of the disk, i.e.\ $PSL(2, \mathbb{R})$ (we are working on the upper half plane again). Of course, this volume is infinite, but we will see that the ratio of this volume and the integral over the position of the vertex operator is finite. This is consistent with the fact that one could alternatively fix the position of the vertex operator, leaving a residual subgroup of the CKG which has finite volume. 

Therefore, we need to calculate
\begin{IEEEeqnarray}{rCl}
\mathcal{A}^\text{closed}_{D_2}(1)&=&g_c\tau_p\int_{\mathbb H_+}\frac{\mathrm d^2z}{V_\text{CKG}}\, \left\langle V^{(1)}(z,\overline z)\right\rangle=g_c\tau_p\int_{\mathbb H_+}\frac{\mathrm d^2z}{V_\text{CKG}}\, \left\langle V^{(1)}(z)\overline V^{(1)}(\overline z)\right\rangle\ .\IEEEeqnarraynumspace
\label{eq::1ptfunction}
\end{IEEEeqnarray}
This seems the obvious guess for the one-point function, but it also comes out from an analysis following \cite{Hoogeveen:2007tu}, as we did for the two-point function above. Plugging in the vertex operator \eqref{V1Mmn}, we obtain
\begin{IEEEeqnarray}{rCl}
\mathcal{A}^\text{closed}_{D_2}(1)&=&g_c\tau_p\int_{\mathbb{H}_+} \frac{\mathrm d^2z}{V_\text{CKG}}\, \langle(\partial X^m A_m[\xi,k](z) - \frac12M^{mn}\mathcal F_{mn}[\xi,k](z))\nonumber\\
&&\hspace{2.3cm} \times(\overline\partial X^r A_r[D{\cdot}\overline\xi,D{\cdot}k](\overline z) - \frac12M^{rs}\mathcal F_{rs}[D{\cdot}\overline\xi,D{\cdot}k](\overline z))\rangle \nonumber \\
&=&g_c\tau_p\int_{\mathbb{H}_+} \frac{\mathrm d^2z}{V_\text{CKG}}\, \langle\partial X^m A_m[\xi,k](z)\overline\partial X^r A_r[D{\cdot}\overline\xi,D{\cdot}k](\overline z)\nonumber\\
&&\hspace{2.3cm} +\frac14M^{mn}\mathcal F_{mn}[\xi,k](z)M^{rs}\mathcal F_{rs}[D{\cdot}\overline\xi,D{\cdot}k](\overline z)\rangle \ .\label{eq::psf_1_pt}
\end{IEEEeqnarray}
Here we have already implicitly used the zero mode prescription to simplify the expression: All terms containing only one Lorentz current  vanish, because they will still contain either $\lambda^\alpha$ or $\theta^\alpha$ after computing the OPEs. The expression in \eqref{eq::psf_1_pt} can now be evaluated by following the same steps as for the two-point function, just using the zero mode prescription $\langle\mathds{1}\rangle_0=1$. We have to compute the OPEs between the left and right moving fields, expand the superfields in $\theta^\alpha$ (the $\theta$-expansions are given in \eqref{expansion}) and discard all terms which contain any power of $\lambda^\alpha$ or $\theta^\alpha$. This results in 
\begin{IEEEeqnarray}{rCl}
\mathcal{A}^\text{closed}_{D_2}(1)&=&g_c\tau_pC_{\mathbb{H}_+}\int_{\mathbb{H}_+} \frac{\mathrm d^2z}{V_\text{CKG}}\, \frac{\left|z-\overline z\right|^{k{\cdot}D{\cdot}k}}{(z-\overline z)^2}\Big(-\eta^{mr}\xi_mD_r^{\hphantom ra}\overline\xi_a+k^aD_a^{\hphantom am}\xi_mk^rD_r^{\hphantom rb}\overline\xi_b\IEEEeqnarraynumspace \nonumber\\ 
&& \hspace{5.25cm} -\eta^{m[s}\eta^{r]n}k_{[m}\xi_{n]}D_{[r}^{\hphantom ra}D_{s]}^{\hphantom sb}k_{a}\overline\xi_{b}\Big) \nonumber\\
&=&g_c\tau_pC_{\mathbb{H}_+}\int_{\mathbb{H}_+} \frac{\mathrm d^2z}{V_\text{CKG}}\, \frac{1}{(z-\overline z)^2}\Big(-\eta^{mr}\xi_mD_r^{\hphantom ra}\overline\xi_a + k^m \xi_m k^b \overline \xi_b - k_{[m}\xi_{n]}k^{m}D^{nb}\overline\xi_{b}\Big)\nonumber\\
&=&-g_c\tau_pC_{\mathbb{H}_+}\int_{\mathbb{H}_+} \frac{\mathrm d^2z}{V_\text{CKG}}\, \frac{1}{(z-\overline z)^2}\text{Tr}(\epsilon{\cdot}D)\ .\label{eq::1ptdiskintermediate}
\end{IEEEeqnarray}
To obtain the first equality in \eqref{eq::1ptdiskintermediate} we have used the OPE of $X^m$ in \eqref{psfOPE} and of the Lorentz current $M^{mn}$ in \eqref{eq::Lorentz_current}. Transversality, masslessness and momentum conservation, 
\begin{IEEEeqnarray}l
(k+D{\cdot}k)^m=0\ ,\label{eq::momentum_cons_1_pt}
\end{IEEEeqnarray}
allowed for further simplification. Moreover, $C_{\mathbb{H}_+}$ is again an overall constant arising from performing the path integral over the non-zero modes of the world-sheet fields. 

What is left, is to calculate the volume of the conformal Killing group on the upper half plane $\mathbb H_+$, i.e.\ $PSL(2,\mathbb R)=SL(2,\mathbb R)/\mathbb Z_2$. Such a transformation is given by 
\begin{IEEEeqnarray}{C}
z\rightarrow\frac{Az+B}{Cz+D}\ , \\
\text{where }
\begin{pmatrix}
A&B\\
C&D
\end{pmatrix}\in SL(2,\mathbb R)\quad\text{i.e.}\quad\left\{\begin{matrix}A,B,C,D\in\mathbb R\ ,\\AD-BC=1\ .\end{matrix}\right.
\end{IEEEeqnarray} 
The discussion is facilitated by making a Cartan decomposition of the group $G=SL(2,\mathbb R)$ \cite{Lang}
\begin{IEEEeqnarray}{l}
\begin{IEEEeqnarraybox}[][c]{rCl}
\IEEEstrut
F:K\times A^+\times K&\longrightarrow&G\\
(k_1,a,k_2)&\longmapsto&g=k_1ak_2\ .
\IEEEstrut
\end{IEEEeqnarraybox}\label{eq::cartan}
\end{IEEEeqnarray}
Here $K$ is the isotropy group of $z=i$, i.e.\ it consists of all $SL(2,\mathbb R)$ transformations with 
\begin{IEEEeqnarray}{l}
k(z=i)=\frac{Ai+B}{Ci+D}=i\ .
\end{IEEEeqnarray}
This leads to 
\begin{IEEEeqnarray}{rCl}
D = A\ , \qquad C = -B\ ,
\label{ADBCrelations}
\end{IEEEeqnarray}
which, together with $AD-BC=1$, implies
\begin{IEEEeqnarray}{rCl}
A^2+B^2=1\ ,\qquad C^2+D^2=1\ .
\label{ABCDconstraints}
\end{IEEEeqnarray}
On the other hand, $A^+$ is the group of all matrices of the from $\begin{pmatrix}a&0\\0&a^{-1}\end{pmatrix}$ with $a>1$. The map $F$ defined in \eqref{eq::cartan} is surjective. However, a Cartan decomposition $g=k_1ak_2$ with $a\in A^+$ and $k_1,k_2\in K$ is determined only up to a factor $\pm1$ in $k_1$ and a related factor in $k_2$ \cite{Lang}. This means that $F$ is of degree 2.

For the calculation of the volume of $SL(2,\mathbb R)$, we need to choose an explicit parametrization of an element $g\in SL(2,\mathbb R)$. In view of \eqref{ADBCrelations} and \eqref{ABCDconstraints} and the definition of $A^+$, this can be chosen as
\begin{IEEEeqnarray}{rCl}
g=\begin{pmatrix}\cos(\theta_1)&\sin(\theta_1)\\-\sin(\theta_1)&\cos(\theta_1)\end{pmatrix}\begin{pmatrix}e^t&0\\0&e^{-t}\end{pmatrix}\begin{pmatrix}\cos(\theta_2)&\sin(\theta_2)\\-\sin(\theta_2)&\cos(\theta_2)\end{pmatrix}\ ,
\label{eq::decomposition}
\end{IEEEeqnarray}
with $t > 0$ and $0\leq\theta_i<2\pi$ for $i=1,2$. With this parametrization a standard measure on $SL(2,\mathbb R)$ is given by (cf.\ section VII, §2 in \cite{Lang})\footnote{The normalization of the measure is actually a matter of convention. A different overall factor can be found, for instance, in formula (4.20) of \cite{Bargmann:1946me}. Different choices for the overall normalization constant are related by different values for $C_{\mathbb{H}_+}$.}
\begin{IEEEeqnarray}{rCl}
\mathrm{d} \mu = \mathrm{d} \theta_1 \mathrm{d} t\, \mathrm{d} \theta_2 \sinh(2t)\ .
\label{eq::measure_sl2R}
\end{IEEEeqnarray}
The volume of $PSL(2,\mathbb R)=SL(2,\mathbb R)/\mathbb Z_2$ is half the volume of $SL(2,\mathbb R)$. Thus, we finally arrive at 
\begin{IEEEeqnarray}{rCl}
V_{CKG}&=&\frac12\int_0^{2\pi}\mathrm d\theta_1\int_0^\infty\mathrm dt\int_0^{2\pi}\mathrm d\theta_2\, \sinh(2t)\ .\label{eq::CKG_hp}
\end{IEEEeqnarray}
In principle, we would have to regularize the volume of the conformal Killing group in \eqref{eq::CKG_hp}, because it is infinite. But actually, it cancels almost (up to a factor $-\pi$) against a similar factor in the one-point function \eqref{eq::1ptdiskintermediate}. This can be seen as follows:
\begin{IEEEeqnarray}{rCl}
\int_{\mathbb{H}_+}\frac{\mathrm d^2z}{(z-\overline z)^2}\overset{z=i\frac{1-z'}{1+z'}}{=}-\int_{D_2}\frac{\mathrm d^2z'}{(1-z'\overline z')^2}\overset{z'=e^{i\varphi}\tanh\left(b\right)}{=}-\int_0^{2\pi}\mathrm d\varphi\int_0^\infty\mathrm db\,\sinh(2b)\ .\IEEEeqnarraynumspace\label{eq::zintegration}
\end{IEEEeqnarray}
Thus, we obtain 
\begin{IEEEeqnarray}{rCl}
\mathcal{A}^\text{closed}_{D_2}(1)&=&g_c\tau_p C_{\mathbb{H}_+} \frac{1}{\pi}\text{Tr}(\epsilon{\cdot}D)\ .
\label{eq::psf1ptdiskresult}
\end{IEEEeqnarray}
This agrees with the corresponding calculation in the RNS formalism, cf.\ appendix \ref{app:1-pt_fct}. 

To determine the constant $C_{\mathbb{H}_+}$ we compare the graviton one-point function obtained from \eqref{eq::psf1ptdiskresult} with the graviton one-point function obtained from the Dirac-Born-Infeld action for the graviton (cf.\ formula (8.7.24) in \cite{Polchinski:1998rq})
\begin{IEEEeqnarray}l
S=\frac{\tau_p}2\int\mathrm d^{p+1}\xi\,h^{m_\parallel}_{\hphantom{m_\parallel}m_\parallel}\ ,
\end{IEEEeqnarray}
where $h_{mn}=G_{mn}-\eta_{mn}=-4\pi g_ce^{ik{\cdot}X}\epsilon_{mn}$  is only traced over the directions tangent to the D$p$-brane, which is denoted by the indices $m_\parallel$.\footnote{In defining $h_{mn}=-4\pi g_ce^{ik{\cdot}X}\epsilon_{mn}$ we follow \cite{Polchinski:1998rq}, cf.\ formula (3.7.11a).} Using Feynman rules we find that 
\begin{IEEEeqnarray}{rCl}
\mathcal A^{graviton}_{D_2}(1)= - 2\pi g_c\tau_p\epsilon^{m_\parallel}_{\hphantom{m_\parallel}m_\parallel}\ .
\label{eq::graviton_DBI}
\end{IEEEeqnarray}
To calculate the graviton amplitude from \eqref{eq::psf1ptdiskresult}, we have to make use of the fact that for a graviton $\text{Tr}(\epsilon)=0$. Hence, we find $\text{Tr}(\epsilon {\cdot} D)=2\epsilon^{m_\parallel}_{\hphantom{m_\parallel}m_\parallel}$. We can conclude that
\begin{IEEEeqnarray}l
\mathcal A^{graviton}_{D_2}(1)= g_c\tau_p C_{\mathbb{H}_+} \frac{2}{\pi }\epsilon^{m_\parallel}_{\hphantom{m_\parallel}m_\parallel}.\label{eq::graviton_purespinor}
\end{IEEEeqnarray}
Comparing the two results in equation \eqref{eq::graviton_DBI} and \eqref{eq::graviton_purespinor}, we can determine the factor $C_{\mathbb{H}_+}=-\pi^2$. Therefore, the final result for a closed string one-point function on the disk is given by
\begin{IEEEeqnarray}{rCl}
\mathcal{A}^\text{closed}_{D_2}(1) = - \pi g_c\tau_p\text{Tr}(\epsilon {\cdot} D)\ .
\label{eq::2ptdiskresult_final}
\end{IEEEeqnarray}

%%%%%%%%%%%%%%%%%%%%%%%%%%%%%%%%%%%%%%

\section{Outlook}

In this paper we showed how to use the pure spinor formalism in order to calculate purely closed string amplitudes on the disk. We focused on the low point functions (i.e.\ the one- and two-point functions) of massless states that would correspond to the NSNS sector in the RNS formalism. A generalization to the massless states of the other RNS sectors would be straightforward. Moreover, the analog of \eqref{eq::2ptdisk} for higher $n$-point functions follows along the same lines, resulting in 
\begin{IEEEeqnarray}{l}
\mathcal{A}^\text{closed}_{D_2}(1,\ldots,n)= \\
= i g_c^n\tau_p\int_0^1\! \mathrm dy \,\biggl\langle \left(V^{(0)}_1(iy)\overline V^{(1)}_1(-iy)+V_1^{(1)}(iy)\overline V^{(0)}_{1}(-iy)\right) \prod_{j=2}^{n-1}\int_{\mathbb{H}_+}\!\! \mathrm d^2z_j V^{(1,1)}_j(z_j,\overline z_j) V^{(0)}_n(i)\overline V^{(0)}_n(-i)\biggl\rangle \nonumber \\ 
= 2 i g_c^n\tau_p\int_0^1\! \mathrm dy \,\biggl\langle V^{(0)}_1(iy)\overline V^{(1)}_1(-iy) \prod_{j=2}^{n-1}\int_{\mathbb{H}_+}\!\! \mathrm d^2z_j V^{(1,1)}_j(z_j,\overline z_j) V^{(0)}_n(i)\overline V^{(0)}_n(-i)\biggl\rangle\ . \nonumber
\label{eq::Nptdisk}
\end{IEEEeqnarray}
In the last equality we used similar BRST arguments as for the two-point function, cf.\  appendix \ref{app:independent}. It would also be interesting to generalize our analysis of the disk to the projective plane. The corresponding two-point function of massless closed strings in the RNS formalism was calculated in \cite{Garousi:2006zh} (see also \cite{Aldi:2020dvw}) and the dilaton one-point function was calculated in \cite{Liu:1987nz,Grinstein:1986hd} for the bosonic string and generalized to type I in \cite{Ohta:1987nq}. 

The results of our paper, taken together with \cite{Mafra:2011nv,Stieberger:2009hq}, should allow us to calculate explicitly higher point functions of closed strings on the disk and in this way learn more about the low energy effective action on the world-volume of D-branes. Optimistically, this might for instance pave the way to verify the existence of an $\epsilon_{10} \epsilon_{10} R^4$-interaction at disk level, predicted in \cite{Green:2016tfs} using heterotic/type I duality. Such a term could have interesting consequences for string theory model building, as it would lead to a disk level correction to the Einstein-Hilbert term in four dimensions after compactification on a manifold with non-vanishing Euler number \cite{Antoniadis:1997eg}. For minimally supersymmetric four dimensional type IIB orientifolds with D9-branes, this would constitute the leading $g_s$-correction to the Einstein-Hilbert term in four dimensions, dominant compared with the one-loop corrections obtained in \cite{Haack:2015pbv}.\footnote{A similar disk level correction to the four dimensional Einstein-Hilbert term was discussed in type IIB orientifolds with D7-branes and O7-planes wrapping four-cycles with non-trivial first Chern form, cf.\ \cite{Weissenbacher:2020cyf}.} Such a correction in the string frame could then lead, via a Weyl rescaling, to a correction to the K\"ahler potential of the moduli and/or to a redefinition of the moduli fields, as discussed for instance in \cite{Berg:2014ama,Haack:2018ufg}.

%%%%%%%%%%%%%%%%%%%%%%%%%%%%%%%%%%%%%%
\vskip 1cm

\noindent
{\Large {\bf Acknowledgments}}

\vskip 3mm

\noindent
We would like to thank Marcus Berg, Carlos Mafra, Ingmar Saberi, Ivo Sachs, Oliver Schlotterer, Dimitri Skliros and Stephan Stieberger for valuable discussions and email correspondence. This work is supported by the Origins Excellence Cluster in Munich.

%%%%%%%%%%%%%%%%%%%%%%%%%%%%%%%%%%%%%%%%%%%%%%%%%%%
 
\appendix
 
\section{Notation}
\label{app:notation}

In this appendix we summarize our notation. We use $\alpha' = 2$ throughout the paper, unless $\alpha'$ is made explicit. We use the following words as synonyms
\begin{IEEEeqnarray*}{rCl}
\text{holomorphic}&=&\text{left moving\ ,}\\
\text{antiholomorphic}&=&\text{right moving\ .}
\end{IEEEeqnarray*} 
Spacetime vector indices are denoted by the small Latin alphabet $a,b,c,\ldots$, while we use small Greek letters $\alpha,\beta,\gamma,\ldots$ for spinor indices. 

In expressions, where we have many indices that are symmetric or antisymmetric, for both vector and spinor indices we use the convention that
\begin{IEEEeqnarray}{rCl}
A^1_{[\alpha_1}A^2_{\alpha_2}\cdots A^n_{\alpha_n]}=\frac{1}{n!}\left(A^1_{\alpha_1}A^2_{\alpha_2}\cdots A^n_{\alpha_n}\pm\text{permutations}\right),\\
A^1_{(\alpha_1}A^2_{\alpha_2}\cdots A^n_{\alpha_n)}=\frac{1}{n!}\left(A^1_{\alpha_1}A^2_{\alpha_2}\cdots A^n_{\alpha_n}+\text{permutations}\right).
\end{IEEEeqnarray}

The Lorentz group in 10 dimensions has two inequivalent representations that are denoted by $16$ and $16'$. Following \cite{Witten:1985nt,Grassi:2003cm} we use the convention (common in the pure spinor literature) that a spinor $\psi^\alpha$ with an upper index transforms in the $16$ representation while a spinor $\psi_\alpha$ with a lower index transforms in the $16'$ representation. As stressed in \cite{Witten:1985nt}, there is no way to raise or lower the indices, as the two representations $16$ and $16'$ are inequivalent. The gamma matrices are symmetric, i.e.\ $\gamma^{m \alpha \beta} = \gamma^{m \beta \alpha}$ and $\gamma^{m}_{\alpha \beta} = \gamma^{m}_{\beta \alpha}$, and they fulfil the following algebra
\begin{IEEEeqnarray}l
\gamma^{m \alpha \epsilon} \gamma^{n}_{\epsilon \beta} + \gamma^{n \alpha \epsilon} \gamma^{m}_{\epsilon \beta}= 2\eta^{mn} \delta^\alpha \, \! _\beta\ ,
\end{IEEEeqnarray}
where $\eta^{mn}$ is the Minkowski metric. In section \ref{sec_Vops}, we use the notation $\gamma^m$ and $\widehat \gamma^m$ when suppressing the lower or upper indices of $\gamma^{m}_{\alpha \beta}$ and $\gamma^{m \alpha \beta}$, respectively. Moreover, for antisymmetric products of gamma matrices we use a similar convention as before so that
\begin{IEEEeqnarray}{rCl}
\gamma^{m_1\ldots m_n} \equiv \gamma^{[m_1\ldots m_n]}=\frac{1}{n!}\left(\gamma^{m_1}\gamma^{m_2}\cdots\gamma^{m_n}\pm\text{permutations}\right).
\end{IEEEeqnarray}
Throughout the text we use round brackets in order to denote contractions of fermions with (products of) gamma matrices, i.e.\
\be
(\psi_1 \gamma^m \psi_2) = \psi_1^\alpha \gamma^m_{\alpha \beta} \psi_2^\beta \qquad {\rm etc.} \ .
\ee
Note that it is implicitly understood here that $\psi_1^\alpha$ is a transposed spinor. 

%%%%%%%%%%%%%%%%%%%%%%%%%%%%%%%%%%%%%%%%%%%%%%%%%%%

\section{Relations for the matrices $M^{\alpha}_{\hphantom \alpha \beta}$ and $N_\alpha^{\hphantom \alpha \beta}$}
\label{app:MN}

In this appendix, we would like to derive the formulas \eqref{NMdelta}-\eqref{eq::gammaup}. The relation \eqref{NMdelta} follows from the OPE
\be
\overline p_\alpha(\overline z) \overline \theta^\beta(\overline w) = N_{\alpha}^{\hphantom \alpha \gamma} p_\gamma (\overline z) M^\beta_{\hphantom \beta \delta} \theta^\delta(\overline w) = N_{\alpha}^{\hphantom \alpha \gamma} M^\beta_{\hphantom \beta \delta}  \frac{\delta_\gamma^{\hphantom\gamma \delta}}{\overline z-\overline w} = \frac{N_{\alpha}^{\hphantom \alpha \gamma} M^\beta_{\hphantom \beta \gamma}}{\overline z-\overline w} \stackrel{!}{=} \frac{\delta_\alpha^{\hphantom\alpha\beta}}{\overline z-\overline w}\ .
\label{conditionptheta}
\ee
The relations of \eqref{eq::gammadown} can be obtained by demanding
\be
\overline \Pi^m(\overline z) \stackrel{!}{=} D^m_{\hphantom mn}\Pi^n(\overline z) \quad , \quad  \overline d_\alpha (\overline z) \stackrel{!}{=} N_\alpha^{\hphantom\alpha \beta} d_\beta (\overline z)\ . \label{Pidtransform}
\ee
We show this exemplarily in the case of the relation involving $M$, as this is the one that we actually use in the main text. From
\be
\overline \Pi^m(\overline z) = \Big( D^m_{\hphantom mn} \overline \partial X^n + \frac12 M^\gamma_{\hphantom \gamma \alpha} \theta^\alpha \gamma^m_{\gamma \delta} M^\delta_{\hphantom \delta \beta} \overline \partial \theta^\beta \Big) (\overline z)  \stackrel{!}{=} D^m_{\hphantom mn}\Pi^n(\overline z) 
\ee
we read off
\be
M^\gamma_{\hphantom \gamma \alpha} \gamma^m_{\gamma \delta} M^\delta_{\hphantom \delta \beta} =  D^m_{\hphantom m n} \gamma^n_{\alpha \beta}\ . \label{MgammadownM}
\ee
It is straightforward to check that \eqref{NMdelta} and \eqref{eq::gammadown} imply consistency of all the OPEs in \eqref{psfOPE} with the doubling trick (i.e.\ the remaining OPEs do not lead to any new conditions). 
 
From the relations \eqref{eq::gammaup} we actually use the one involving $N$ in the main text. That relation follows from demanding
\be
\overline \Sigma^{mn} (\overline z) \stackrel{!}{=} D^m_{\hphantom m k} D^n_{\hphantom n l} \Sigma^{kl} (\overline z)\ .
\ee
This holds if in addition to \eqref{MgammadownM} we have 
\be
N_\gamma^{\hphantom \gamma \alpha} \gamma^{m \gamma \delta} N_\delta^{\hphantom \delta \beta} = D^m_{\hphantom m n} \gamma^{n \alpha \beta} \label{NgammaupN}
\ee
because
\be
\overline \Sigma^{mn} (\overline z) & = &  - \frac12 (\overline p \gamma^{mn} \overline \theta) (\overline z) \\
& = & - \frac14 \overline p_\alpha (\gamma^{m \alpha \delta} \gamma^n_{\delta \beta} - \gamma^{n \alpha \delta} \gamma^m_{\delta \beta}) \overline \theta^\beta (\overline z) \\
& = & - \frac14 N_\alpha^{\hphantom \alpha \epsilon} p_\epsilon (\gamma^{m \alpha \delta} \gamma^n_{\delta \beta} - \gamma^{n \alpha \delta} \gamma^m_{\delta \beta}) M^\beta_{\hphantom \beta \rho} \theta^\rho (\overline z) \\
& = &  - \frac14 p_\epsilon \theta^\rho \Big( N_\alpha^{\hphantom \alpha \epsilon} \gamma^{m \alpha \sigma} N_\sigma^{\hphantom \sigma \gamma} M^\xi_{\hphantom \xi \gamma} \gamma^n_{\xi \beta} M^\beta_{\hphantom \beta \rho} - N_\alpha^{\hphantom \alpha \epsilon} \gamma^{n \alpha \sigma} N_\sigma^{\hphantom \sigma \gamma} M^\xi_{\hphantom \xi \gamma} \gamma^m_{\xi \beta} M^\beta_{\hphantom \beta \rho} \Big)(\overline z) \label{barSigma4} \\
& = & - \frac14 p_\epsilon \theta^\rho \Big( D^m_{\hphantom m k} \gamma^{k \epsilon \gamma} D^n_{\hphantom n l} \gamma^l_{\gamma \rho}  - D^n_{\hphantom m l} \gamma^{l \epsilon \gamma} D^m_{\hphantom n k} \gamma^k_{\gamma \rho} \Big)(\overline z) \label{barSigma5} \\
& = & - \frac14 D^m_{\hphantom m k} D^n_{\hphantom n l} p_\epsilon (\gamma^{k \epsilon \gamma} \gamma^l_{\gamma \rho} - \gamma^{l \epsilon \gamma} \gamma^k_{\gamma \rho}) \theta^\rho (\overline z) \\
& = & D^m_{\hphantom m k} D^n_{\hphantom n l} \Sigma^{kl} (\overline z)\ .
\ee
In \eqref{barSigma4} we used \eqref{NMdelta} and in \eqref{barSigma5} we employed \eqref{MgammadownM} and \eqref{NgammaupN}.
%%%%%%%%%%%%%%%%%%%%%%%%%%%%%%%%%%%%%%%%%%%%%%%%%%%

\section{Independence of the correlator of the localization of the integrated vertex operator}
\label{app:independent}

If we make the dependence of the vertex operator on the polarization vector and the momentum explicit, we can write \eqref{eq::redefinedvertex1} and \eqref{eq::redefinedvertex2} more explicitly as
\begin{IEEEeqnarray}{rCl}
\overline V^{(0)}[\overline\xi,k](\overline z)&=& V^{(0)}[D {\cdot} \overline\xi, D {\cdot} k](\overline z)\ ,\\
\overline V^{(1)}[\overline\xi,k](\overline z)&=& V^{(1)}[D {\cdot} \overline\xi, D {\cdot} k](\overline z)\ .
\end{IEEEeqnarray}
Using this, we can rewrite the first summand of \eqref{eq::2ptdisk} according to 
\begin{IEEEeqnarray}{rCl}
& & \int_0^1\mathrm dy\,\biggl\langle V^{(0)}_1[\xi_1,k_1] (iy) \ \overline V^{(1)}_1[\overline \xi_1,k_1] (-iy) \ V^{(0)}_2[\xi_2,k_2](i)\ \overline V^{(0)}_2[\overline \xi_2, k_2](-i)\biggl\rangle \nonumber \\
& = & \int_0^1\mathrm dy\,\biggl\langle V^{(1)}_1[\xi_1,k_1] (iy) \ \overline V^{(0)}_1[\overline \xi_1,k_1] (-iy) \ V^{(0)}_2[\xi_2,k_2](i)\ \overline V^{(0)}_2[\overline \xi_2, k_2](-i)\biggl\rangle \ ,
\label{eq::2ptdisk_summands}
\end{IEEEeqnarray}
i.e.\ the correlator is independent of whether the integrated vertex operator is located at $i y$ or at $-i y$. This can be shown explicitly using Cadabra, or it can be seen as follows:
\begin{IEEEeqnarray}{l}
\int_0^1\mathrm dy\,\left\langle V^{(0)}_1[\xi_1,k_1](iy)\overline V^{(1)}_1[\overline\xi_1,k_1](-iy)V^{(0)}_2[\xi_2,k_2](i)\overline V^{(0)}_2[\overline\xi_2,k_2](-i)\right\rangle \nonumber \\
=\int_0^1\mathrm dy\,\left\langle V^{(0)}_1[\xi_1,k_1](iy)V^{(1)}_1[D{\cdot}\overline\xi_1,D{\cdot}k_1](-iy)V^{(0)}_2[\xi_2,k_2](i)V^{(0)}_2[D{\cdot}\overline\xi_2,D{\cdot}k_2](-i)\right\rangle  \nonumber \\
=\int_0^1\mathrm dy\,\left\langle \int_{-iy}^{iy}\mathrm dz\,\partial V^{(0)}_1[\xi_1,k_1](z)V^{(1)}[D{\cdot}\overline\xi_1,D{\cdot}k_1]_1(-iy)V^{(0)}_2[\xi_2,k_2](i)V^{(0)}_2[D{\cdot}\overline\xi_2,D{\cdot}k_2](-i)\right\rangle \label{independence_1} \\
=\int_0^1\mathrm dy\int_{-iy}^{iy}\mathrm dz\,\left\langle QV^{(1)}_1[\xi_1,k_1](z)V^{(1)}_1[D{\cdot}\overline\xi_1,D{\cdot}k_1](-iy)V^{(0)}_2[\xi_2,k_2](i)V^{(0)}_2[D{\cdot}\overline\xi_2,D{\cdot}k_2](-i)\right\rangle \label{independence_2} \\
=-\int_0^1\mathrm dy\int_{-iy}^{iy}\mathrm dz\,\left\langle V^{(1)}_1[\xi_1,k_1](z)QV^{(1)}_1[D{\cdot}\overline\xi_1,D{\cdot}k_1](-iy)V^{(0)}_2[\xi_2,k_2](i)V^{(0)}_2[D{\cdot}\overline\xi_2,D{\cdot}k_2](-i)\right\rangle  \label{independence_3} \\
=\int_0^1\mathrm dy\int_{-iy}^{iy}\mathrm dz\,\left\langle V^{(1)}_1[\xi_1,k_1](z)i\partial_yV^{(0)}_1[D{\cdot}\overline\xi_1,D{\cdot}k_1](-iy)V^{(0)}_2[\xi_2,k_2](i)V^{(0)}_2[D{\cdot}\overline\xi_2,D{\cdot}k_2](-i)\right\rangle \label{independence_4} \\
=i\int_0^1\mathrm dy\,\partial_y\left\langle\int_{-iy}^{iy}\mathrm dz\, V^{(1)}_1[\xi_1,k_1](z)V^{(0)}_1[D{\cdot}\overline\xi_1,D{\cdot}k_1](-iy)V^{(0)}_2[\xi_2,k_2](i)V^{(0)}_2[D{\cdot}\overline\xi_2,D{\cdot}k_2](-i)\right\rangle \nonumber \\
\hphantom=-i\int_0^1\mathrm dy\,\left\langle \Big( \partial_y\int_{-iy}^{iy}\mathrm dz V^{(1)}_1[\xi_1,k_1](z) \Big)V^{(0)}_1[D{\cdot}\overline\xi_1,D{\cdot}k_1](-iy)V^{(0)}_2[\xi_2,k_2](i)V^{(0)}_2[D{\cdot}\overline\xi_2,D{\cdot}k_2](-i)\right\rangle \label{independence_5} \\
= i \int_{-i}^{i}\mathrm dz\,\left\langle V^{(1)}_1[\xi_1,k_1](z)V^{(0)}_1[D{\cdot}\overline\xi_1,D{\cdot}k_1](-i)V^{(0)}_2[\xi_2,k_2](i)V^{(0)}_2[D{\cdot}\overline\xi_2,D{\cdot}k_2](-i)\right\rangle \nonumber \\
\hphantom{=}-i\int_{0}^{0}\mathrm dz\,\left\langle V^{(1)}_1[\xi_1,k_1](z)V^{(0)}_1[D{\cdot}\overline\xi_1,D{\cdot}k_1](0)V^{(0)}_2[\xi_2,k_2](i)V^{(0)}_2[D{\cdot}\overline\xi_2,D{\cdot}k_2](-i)\right\rangle \label{independence_6} \\
\hphantom{=} + \int_0^1\mathrm dy\left\langle \Big(V^{(1)}_1[\xi_1,k_1](iy)+V^{(1)}_1[\xi_1,k_1](-iy) \Big) V^{(0)}_1[D{\cdot}\overline\xi_1,D{\cdot}k_1](-iy)V^{(0)}_2[\xi_2,k_2](i)V^{(0)}_2[D{\cdot}\overline\xi_2,D{\cdot}k_2](-i)\right\rangle \nonumber \\
=\int_0^1\mathrm dy\left\langle V^{(1)}_1[\xi_1,k_1](iy) V^{(0)}_1[D{\cdot}\overline\xi_1,D{\cdot}k_1](-iy)V^{(0)}_2[\xi_2,k_2](i)V^{(0)}_2[D{\cdot}\overline\xi_2,D{\cdot}k_2](-i)\right\rangle\ .  \label{independence_7}
\end{IEEEeqnarray}
In \eqref{independence_1} we rewrote $V^{(0)}_1[\xi_1,k_1]$ as an integral over $\partial V^{(0)}_1[\xi_1,k_1]$ and used that there is no contribution from the lower integration end due to the ``cancelled propagator argument'', which states that terms with vertex operators at the same position vanish, cf.\ p.\ 196 in \cite{Polchinski:1998rq}. In \eqref{independence_2} and \eqref{independence_4} we used \eqref{QV1}, in \eqref{independence_3} we deformed the BRST contour and used \eqref{QV0}, in \eqref{independence_5} we performed a partial integration and in \eqref{independence_7} we again used the cancelled propagator argument. 

The analysis above can be generalized to an $n$-point function with $n>2$, i.e.\ to the case when there are additional integrated closed string vertex operators in the correlation function, cf.\ \eqref{eq::Nptdisk}. When deforming the BRST contour in that case, the BRST charge $Q$ also acts on the integrated vertex operators $V_j^{(1,1)}$ for $j=2,\ldots,n-1$. However, this does not give any additional contributions, as can be seen as follows:
\be
&& \hspace{-2cm} Q \int_{\mathbb{H}^+} d^2 z V^{(1)} [\xi,k](z) V^{(1)}[D {\cdot} \overline \xi, D {\cdot} k](\overline z) \nonumber \\ 
& = & \int_{\mathbb{H}^+} d^2 z \Big( \partial V^{(0)} [\xi,k](z) V^{(1)}[D {\cdot} \overline \xi, D {\cdot} k](\overline z) + V^{(1)} [\xi,k](z) \overline \partial V^{(0)}[D {\cdot} \overline \xi, D {\cdot} k](\overline z) \Big) \nonumber \\ 
& = & \int_{\mathbb{H}^+} d^2 z \Big( \partial \Big[ V^{(0)} [\xi,k](z) V^{(1)}[D {\cdot} \overline \xi, D {\cdot} k](\overline z) \Big] + \overline \partial \Big[ V^{(1)} [\xi,k](z)  V^{(0)}[D {\cdot} \overline \xi, D {\cdot} k](\overline z) \Big] \Big) \nonumber \\ 
& \sim & \int_{\mathbb{R}} dx \Big( V^{(0)} [\xi,k](x) V^{(1)}[D {\cdot} \overline \xi, D {\cdot} k](x) + V^{(1)} [\xi,k](x)  V^{(0)}[D {\cdot} \overline \xi, D {\cdot} k](x) \Big) \label{divergence_theorem} \\
& \rightarrow & 0\ .
\ee
We used the divergence theorem in \eqref{divergence_theorem} (in the form of (2.1.9) in \cite{Polchinski:1998rq}) and the final vanishing result holds inside the correlator when applying the cancelled propagator argument again. 

%%%%%%%%%%%%%%%%%%%%%%%%%%%%%%%%%%%%%%%%%%%%%%%%%%%

\section{NSNS one-point function on the disk in the RNS formalism}
\label{app:1-pt_fct}

For ease of comparison, in this appendix we include the calculation of the NSNS one-point function on the disk in the RNS formalism, even though it is not new. As usual, we could either map the one-point function from the disk to the upper half plane or calculate it directly on the disk. We will perform the calculation on the upper half plane in this appendix, as we did for the pure spinor formalism in the main text.

For the bosonic string the one-point function of a closed string on the disk was calculated in \cite{Douglas:1986eu} and a generalization for the superstring can be found in \cite{Ohta:1987nq}.

On the upper half plane we need a total superghost charge $q=-2$. Hence, the vertex operator has to be in the $(-1,-1)$ picture. Therefore,we need to calculate
\begin{IEEEeqnarray}{rCl}
\mathcal{A}^\text{closed}_{D_2}(1)=g_c\tau_p\int\frac{\mathrm d^2z}{V_\text{CKG}}\, \langle V_{(-1,-1)}(z,\overline z)\rangle\ .\label{eq::1pt_disk_start}
\end{IEEEeqnarray}
As in \eqref{eq::1ptfunction} we did not fix the conformal Killing group symmetry (hence there are no $c$-ghost insertions). Instead we kept the factor $\frac{1}{V_\text{CKG}}$ explicitly and we will show that it cancels against a similar factor in the amplitude, leaving a finite result. Again we can split the vertex operator into a holomorphic and an antiholomorphic part $V(z,\overline z)=\epsilon_{m n}V^m_{-1}(z)\overline V^n_{-1}(\overline z)$, where
\begin{IEEEeqnarray}{l}
V^m_{-1}(z)=e^{-\phi(z)}\psi^m(z)e^{ik{\cdot}X(z)}
\label{eq::vertex_picture}
\end{IEEEeqnarray}
and the antiholomorophic part can be determined via the doubling trick. Concretely, we can extend the fields to the entire complex plane in the following way
\begin{IEEEeqnarray}{rCl}
X^m(z)&=&\left\{\begin{array}{r l}X^m(z)&\text{for }z\in\mathbb H^+\\D^m_{\hphantom{m}n}\overline X^n(z)&\text{for }z\in\mathbb H^-\end{array}\right.,\\
\psi^m(z)&=&\left\{\begin{array}{r l}\psi^m(z)&\text{for }z\in\mathbb H^+\\D^m_{\hphantom{m}n}\overline\psi^n(z)&\text{for }z\in\mathbb H^-\end{array}\right..
\end{IEEEeqnarray}
The matrix $D^{mn}$ was introduced in \eqref{eq::boundary_matrix}. It accounts for the boundary conditions. This allows us then to make the replacement
\begin{IEEEeqnarray}{l}
\overline X^m(\overline z)\longrightarrow D^m_{\hphantom{m}n}X^n(\overline z),\qquad\overline\psi^m(\overline z)\longrightarrow D^m_{\hphantom{m}n}\psi^n(\overline z),\qquad\overline\phi(\overline z)\longrightarrow\phi(\overline z)\ . 
\label{eq::RNS_replace}
\end{IEEEeqnarray}
With this replacement we can express the vertex operator solely by \eqref{eq::vertex_picture}:
\begin{IEEEeqnarray}{rCl}
\begin{IEEEeqnarraybox}[][c]{rCl}
\IEEEstrut
V_1(z_1,\overline z_1)&=&\epsilon_{1mn}D^n_{\hphantom{n}r}V^m_{-1}(k_1,z_1)V_{-1}^r(D{\cdot}k_1,\overline z_1)\ .
\IEEEstrut
\end{IEEEeqnarraybox}\label{eq::vertex_operator_RNS_2}
\end{IEEEeqnarray}

For the calculation on the upper half plane, we need the correlators of the world-sheet fields on $\mathbb H_+$, which are given by
\begin{IEEEeqnarray}{rCl}
\begin{IEEEeqnarraybox}[][c]{rCl}
\IEEEstrut
\langle X^m(z)\overline X^n(\overline w)\rangle&=&-D^{mn}\ln(z-\overline w)\ ,\\
\langle \psi^m(z)\overline\psi^n(\overline w)\rangle&=& \frac{D^{mn}}{z-\overline w}\ ,\\
\langle e^{-\phi(z)} e^{-\overline \phi(\overline w)} \rangle&=& \frac{1}{z-\overline w}\ .
\IEEEstrut
\end{IEEEeqnarraybox}\label{eq::disk correlator}
\end{IEEEeqnarray}
With these correlators we can evaluate 
\begin{IEEEeqnarray}{rCl}
\mathcal{A}^\text{closed}_{D_2}(1)&=&g_c\tau_p\int\frac{\mathrm d^2z}{V_\text{CKG}}\, \,\langle\epsilon_{mn}e^{-\phi(z)}\psi^m(z)e^{ik{\cdot}X(z)}e^{-\overline\phi(\overline z)}\overline\psi^n(\overline z)e^{ik{\cdot}\overline X(\overline z)} \rangle
\label{eq::startingpoint}
\end{IEEEeqnarray}
to be
\begin{IEEEeqnarray}{rCl}
\mathcal{A}^\text{closed}_{D_2}(1)&=& g_c\tau_p \tilde C_{\mathbb{H}_+}\int_{\mathbb H_+}\frac{\mathrm d^2z}{V_\text{CKG}}\, \frac{\text{Tr}(\epsilon {\cdot} D)}{(z-\overline z)^2}\ .
\label{eq::intermediate_result}
\end{IEEEeqnarray}
The Koba-Nielson factor $\left|z - \overline z\right|^{k{\cdot}D{\cdot}k}$ is equal to one, because $k{\cdot}D{\cdot}k=-k^2=0$ due to momentum conservation \eqref{eq::momentum_cons_1_pt} and the fact that we are looking at massless states. The factor $\tilde C_{\mathbb{H}_+}$ accounts for the functional determinants of the world-sheet fields, like in the pure spinor formalism. Of course, there is no reason why $\tilde C_{\mathbb{H}_+}$ should be equal to the corresponding constant $C_{\mathbb{H}_+}$ in the pure spinor formalism. And indeed, \eqref{eq::intermediate_result} agrees with \eqref{eq::1ptdiskintermediate} only if $\tilde C_{\mathbb{H}_+} = - C_{\mathbb{H}_+}$. Assuming this, the two results are identical and, thus, \eqref{eq::intermediate_result} can be treated in exactly the same way as in section \ref{1pt}.  

%%%%%%%%%%%%%%%%%%%%%%%%%%%%%%%%%%%%%%%%%%%%%%%%%%%

\end{document}